\documentclass[a4paper,11pt]{article}
\pdfoutput=1
\usepackage{jheppub}
\usepackage{epsf}
\usepackage{youngtab}
\usepackage{multirow}
\usepackage{dcolumn}
\usepackage{bm}
\usepackage{slashed}
\usepackage{hyperref}
\usepackage{dsfont}
\usepackage{booktabs}
\usepackage{cleveref}
\usepackage{ulem}
\usepackage{titlesec}

\renewcommand{\tilde}{\widetilde} 
\renewcommand{\hat}{\widehat}

\newcommand{\beq}{\begin{eqnarray}}
\newcommand{\eeq}{\end{eqnarray}}

\newcommand{\br}{\mathrm{Br}}

\newcommand{\eps}{\epsilon}

\newcommand{\GeV}{\,\mathrm{GeV}}
\newcommand{\TeV}{\,\mathrm{TeV}}

\newcommand{\fb}{\,\mathrm{fb}}
\newcommand{\MET}{{\slashed{E}_T}}

\begin{document}

\title{LHC vector resonance searches in the $t\bar{t} Z$ final state}
\date{\today}
\preprint{CP3-16-53, CTPU-16-30, KIAS-P16078}

\author[a]{Mihailo~Backovi\'c,}
\author[b,c]{Thomas~Flacke,}
\author[d]{Bithika Jain,}
\author[c,d]{Seung~J.~Lee}


\affiliation[a]{Center for Cosmology, Particle Physics and Phenomenology - CP3,\\
Universite Catholique de Louvain, Louvain-la-neuve, Belgium}
\affiliation[b]{Center for Theoretical Physics of the Universe, IBS, Daejeon, Korea}
\affiliation[c]{Department of Physics, Korea University, Seoul 02841, Korea}
\affiliation[d]{School of Physics, Korea Institute for Advanced Study, Seoul 02455, Korea}

\emailAdd{mihailo.backovic@uclouvain.be}
\emailAdd{flacke@ibs.re.kr }
\emailAdd{ bjain@kias.re.kr}
\emailAdd{sjjlee@korea.edu}

\abstract{
LHC searches for BSM resonances in $l^+ l^-, \, jj, \, t\bar{t}, \gamma \gamma$ and $VV$ final states have so far not resulted in discovery of new physics. Current results set lower limits on mass scales of new physics resonances well into the $\mathcal{O}(1)$ TeV range, assuming that the new resonance decays dominantly to a pair of Standard Model particles. While the SM pair searches are a vital probe of possible new physics, it is important to re-examine the scope of new physics scenarios probed with such final states. Scenarios where new resonances decay dominantly to final states other than SM pairs, even though well theoretically motivated, lie beyond the scope of SM pair searches. 
In this paper we argue that LHC searches for (vector) resonances beyond two particle final states would be useful complementary probes of new physics scenarios. 
As an example, we consider a class of composite Higgs models, and identify specific model parameter points where the color singlet, electrically neutral vector resonance $\rho_0$ decays dominantly not to a pair of SM particles, but to a fermionic top partner $T_{f1}$ and a top quark, with $T_{f1} \rightarrow tZ$. We show that dominant decays of $\rho_0 \rightarrow T_{f1} t$ in the context of Composite Higgs models are possible even when the decay channel to a pair of $T_{f1}$ is kinematically open. Our analysis deals with scenarios where both $m_\rho$ and $m_{T_{f1}}$ are of $\mathcal{O}(1)$ TeV, leading to highly boosted $t\bar{t}Z$ final state topologies. We show that the particular composite Higgs scenario we consider is discoverable at the LHC13 with as little as $30 \fb^{-1}$, while being allowed by other existing experimental constraints.

}

\maketitle


\section{Introduction}\label{sec:intro}


Searches for Beyond the Standard Model (BSM) vector, scalar and fermionic resonances constitutes a significant portion of the ATLAS and CMS exotica program. Past efforts have largely (but not exclusively) focused on searches for statistically significant ``bumps'' in the invariant mass spectra of two final state leptons, jets, top quarks, photons or $W$ and $Z$ bosons.
The apparent absence of an excess in the Standard Model (SM) pair searches pushed the lower mass limits on the new vector resonances well into the  $O(1) \TeV$ range, in models where the branching ratios (BR) to Standard Model (SM) pairs dominates ~\cite{ATLAS:2016ICHEP,CMS:2016ICHEP}. It is certainly interesting to entertain the idea that if new resonances exist, they could (in principle) dominantly decay to final states more complicated than SM pairs. The branching ratios of the new resonance could be such that the signal yield into the SM pairs is too small to be observed at the LHC with the current amount data. 

Resonance decay modes which are more complex than a pair of SM particles are present in many interesting models. For example, some phenomenologically viable models motivated by flat~\cite{Cheng:1999bg} or warped~\cite{Randall:1999ee} extra dimension
where a color charged resonance such as a Kaluza Klein (KK) gluon could decay into a top and a vector like top partner ($T'$). Such models have been studied in much detail in Ref.~\cite{Dobrescu:2009vz} in the context of Tevatron physics and later, in light of early LHC physics~\cite{Kong:2011aa,Contino:2006nn,Contino:2008hi,Bini:2011zb,Barcelo:2011wu}. Furthermore, Refs.~\cite{Agashe:2007ki,Redi:2013eaa, Vignaroli:2014bpa, Chala:2014mma,Greco:2014aza,Barducci:2015vyf} discussed searches for electroweak spin-1 resonances decaying into a pair of top partners as well as into a third generation quark and a top partner, and showed that the current limits on vector resonance masses could be significantly weakened if the resonance decays dominantly to final states other than SM pairs. 

Another interesting example can be found in Composite Higgs Models (CHMs)~\cite{Dugan:1984hq,Georgi:1974yw,Kaplan:1983fs,Kaplan:1983sm,Contino:2003ve,Contino:2010rs,Rychkov:2011br,Panico:2015jxa,Agashe:2004rs,Gripaios:2009pe,Mrazek:2011iu}, where several phenomenologically viable scenarios have been proposed in the past~\cite{Matsedonskyi:2012ym,Redi:2012ha,Marzocca:2012zn,Pomarol:2012qf,Panico:2012uw}. In CHMs, the Higgs typically appears as a pseudo Nambu-Goldstone Boson (pNGB) of an extended global symmetry~\cite{Agashe:2004rs,Contino:2006qr,Contino:2010rs}, 
and can be significantly lighter than the compositeness scale, $f$, in a natural way. The ``composite sector'' of the broken extended gauge symmetry yields additional  bound states apart from the Higgs scalar, for example analogues of the $\rho$ mesons of  QCD.  The new states are typically expected to appear around the mass scale  $\Lambda=4\pi f$, in order to ensure the unitarity in $WW$ scattering~\cite{LlewellynSmith:1973yud,Dicus:1992vj,Cornwall:1973tb,Cornwall:1974km,Lee:1977yc,Lee:1977eg,Chanowitz:1985hj}. Explicit CHMs also necessitate existence of  ``top partners", fermionic resonances required to cancel the divergent contributions from the top quarks with masses, $m_{\Psi}\sim f$.

Electroweak precision (EWP) measurements from collider experiments ~\cite{Barbieri:2012tu, Grojean:2013qca,Ciuchini:2013pca,deBlas:2016ojx} provide indirect constraints on the new gauge bosons which can mix with the $W$ and $Z$.  In CHMs, the most stringent limit arises from the oblique parameter $\hat{S}$  which yields bounds which are typically around 2 TeV~\cite{Panico:2015jxa,Barducci:2015vyf,Contino:2015mha}. Furthermore,  direct searches for top partners at the LHC~\cite{ATLAS:2016ICHEP,CMS:2016ICHEP} put a lower limit on the top partner masses of  $\sim950$ GeV~\footnote{For phenomenological studies see Refs.~\cite{Backovic:2015bca, Backovic:2015lfa, Backovic:2014uma, Azatov:2013hya,Reuter:2014iya,Matsedonskyi:2014mna,Gripaios:2014pqa,Ortiz:2014iza,Han:2014qia}.}.


In this paper, we study in detail the phenomenology of a CHM model with an electroweak neutral vector resonance , $\rho_0$ (sometimes called $Z'$), in the part of the parameter space where $\rho_0$ decays dominantly to a top quark and its partner (for standard SM searches see Ref.~\cite{Low:2015uha,Agashe:2007ki}).  The top partner is referred as $T'$ in general,  however in the  setup we use we label it as the $T_{f1}$.  The model constitutes an example of a non-traditional vector resonance search in the context of a \textit{electroweak neutral} resonance (also see Ref.~\cite{Greco:2014aza}), as well as a feasibility study for resonance searches in the $t\bar{t}Z$ final state at the LHC Run II. The model also represents an example of a CHM where the decays $\rho \rightarrow t\, T_{f1}$ can dominate even when the $\rho_0 \rightarrow T_{f1} T_{f1}$ channel is kinematically open. We identify specific points in the model parameter space which give dominant decays of $\rho_0 \rightarrow tT_{f1}, $ and are explicitly not ruled out by existing experimental results.  However, it is important to note that our main interest here is to explore the novelties of the resonant $t\bar{t}Z$ final state, rather than signatures of particular models, as the resonant $t\bar{t}Z$ final state could be important for a broad range of New Physics  (NP) scenarios~\footnote{For a recent first study by CMS for $Z' \to T' t \to t t Z$ see Ref.~\cite{CMS:2016usi}. }. Considering for instance a  color octet vector resonance instead of a color singlet (electroweak neutral) would change the resonance production mechanism, potentially allowing higher production cross sections by the current collider constraints~\cite{Chala:2014mma}.  We hence consider our results for the CHM model to be a conservative, but theoretically well motivated example of a possible new physics search in the $t\bar{t}Z$ final state. 

Scenarios where $\rho_0$ resonance decays dominantly into a top quark and a top partner, which subsequently decays into a top quark and a $Z$ or an $h$ offer rich LHC phenomenology.  The main signature of the new vector resonance is a final state consisting of three (reconstructed) heavy SM particles, with a distinct ``double resonant'' structure in the invariant mass spectra of two and three reconstructed heavy SM particles. Throughout the paper we consider $\rho_0$ with mass $\sim 2 \TeV$ and the top partner of mass $\sim 1 \TeV$, leading to event topologies where are heavy SM objects in the $t\bar{t}Z$ final state are highly boosted. The highly boosted regime offers both good prospects for accurate reconstruction of the $\rho_0$ and $T_{f_1}$ resonance masses, as well as less contamination from SM backgrounds which rapidly fall of with the increase in transverse momenta. Note that searches for resonances in  three body decays have so far been overlooked at the LHC, despite much top-down motivation to examine such final states~\cite{Contino:2006nn,Greco:2014aza,Barducci:2015vyf,Agashe:2007ki}. 

We study two concrete phenomenological scenarios: one in which the  two high $p_T$ final state top quarks decay hadronically (into jets with substructure, referred to as ``fat jets'') while the $Z$ boson decays into leptons, and the other, where one of the tops decays leptonically, while the other top and the $Z$ decay into hadrons. In the specific example of the CHM, our results show that the two fat jet + two lepton final state can provide enough sensitivity to discover a new color single vector resonance with as little as $\sim 30 \fb^{-1}$ in the $t\bar{t}Z$ final state, while the SM $t\bar{t}$+jets background proves to be a significant challenge for a search in the 2 fat jet + lepton + missing energy channel. We also point out that searches for $\rho \rightarrow T_{f1} t$  processes can benefit from the characteristic resonant structure which appears in the Dalitz-like planes of invariant masses of two and three reconstructed heavy SM objects. 

In section~\ref{sec:models} of the paper we introduce the benchmark CHM model we use throughout the paper, discuss the existing constraints and identify three viable benchmark points. In section~\ref{sec:pheno} we discuss the collider phenomenology and the ability of the LHC to probe the doubly resonant $ttZ$ final states. We give our summary and discussions in section~\ref{sec:concl}, while we present further details of the benchmark model in the Appendix~\ref{app:1} and~\ref{app:2}.

\section{A sample model: the two-site Composite Higgs Model}\label{sec:models}


As a sample model for our study, we use the two site composite Higgs scenario~\cite{Panico:2011pw} and follow the notations and conventions used in Ref.~\cite{Low:2015uha}. The two site model is a simplified version of the five-dimensional (5D) model with $SO(5)\rightarrow SO(4)$ breaking, based on dimensional deconstruction.  As compared to a full 5D model  (or a 4D dual composite model with a full spectrum of resonances) it has a much simpler mass spectrum as it contains only the lowest Kaluza Klein (KK) resonances. 
The two site model contains only the zero level modes (to be identified with the Standard Model particles) and the first KK modes. Refs. ~\cite{Panico:2011pw,Matsedonskyi:2012ym} showed that adding just one set of heavy resonances arising from strong dynamics is sufficient to give a  leading order finite Higgs potential and the electroweak precisions variables i.e $\hat{S}$ and $\hat{T}$.
The model thus provides a ultra-violet (UV) stable setup 
which is sufficient to study the ``low-energy" dynamics of the heavy vector and fermionic resonances.  

In the next subsection, we briefly review the particle content, the mass spectrum, and the interactions (at least those relevant for this article) of the two site model. More details on the model can be found in Appendix~\ref{app:1}. In subsection \ref{subsection:constraints} we summarize current constraints on the model from electroweak precision tests, flavor physics, modifications of Higgs and $Z$ couplings, as well as constraints from resonance searches at LHC run I and II. In subsection \ref{sec:benchmark}, we define 3 benchmark points which are not in contradiction with the current constraints, and are used for our detailed phenomenology study of a search for an electroweak resonance in the $t\bar{t}Z$ final state. 

\subsection{Symmetries and particle content of the two site composite Higgs model}\label{subsec:2sitereview}

The two site composite Higgs model is a non-linear $\sigma$-model with a global symmetry $SO(5)_L\times SO(5)_R$  which is spontaneously broken to its  diagonal group $SO(5)_V$. The  corresponding Goldstone matrix, $\mathcal{U}$, of $SO(5)_L\times SO(5)_R / SO(5)_V$ contains 10 Goldstone bosons in the adjoint of $SO(5)_V$, which decompose into a $\bf{4}$ and a $\bf{6}$ under $SO(4)_V\subset SO(5)_V$. The $\bf{4}$ will be identified with the composite Higgs while the $\bf{6}$ will provide longitudinal degrees of freedom for the heavy gauge bosons. 

Let us first look at the gauge sector of the model. The $SO(5)_L$ site (first site) is associated to the elementary sector while the second site is related to the composite sector. To ensure the correct hyper-charges for the fermions, an extra $U(1)_X$ is added on both sites.  Elementary gauge bosons are added to the first site by gauging the $SU(2)_L\times U(1)_Y$ subgroup of $SO(5)\times U(1)_X$, yielding gauge fields $\hat{W}$ and $\hat{B}$  with couplings $\hat{g}$ and $\hat{g}'$.  To get heavy vector resonances, an  $SO(4) \subset SO(5)_R$ in the second site is gauged with a strong coupling $g_{\rho}$.

The electroweak SM gauge fields are a linear combination of the elementary group $SU(2)_L \times  U(1)_Y$  inside $SO(5)_L$  and the analogous subgroup residing in the $SO(5)_R$ . This combination belongs to the $SO(5)_V$  symmetry group and is unbroken before the Higgs takes a VEV. The heavy vector bosons form a $\bf{6}$ under $SO(4)_V$ which decomposes into $\mathbf{3}_0+ \mathbf{1}_0+\mathbf{1}_{\pm}$ under $SU(2)_L\times U(1)_Y$.  We denote the $SU(2)_L$ triplet vector states with $\rho_{0,\pm}$ and the $SU(2)_L$ singlets with $\rho^0_B$ and $\rho^{\pm}_C$.  The heavy gauge bosons acquire couplings to to SM gauge bosons upon electro-weak symmetry breaking. \footnote[2]{The coupling is present even in the absence of electroweak symmetry breaking. Once the Higgs vacuum expectation $v$ breaks electroweak symmetry, the heavy gauge bosons and the EW gauge bosons obtain additional mass mixing contributions of the order $v/f$.} 
Note that  our phenomenological studies in the following sections  will focus on the neutral state in the $SU(2)_L$ triplet, $\rho_0$.

We proceed with the discussion of  fermion couplings. The Higgs potential is mostly affected by the dynamics associated with the top quark and the composite states it mixes with, i.e. the top partners, while other quark partners typically have less of an influence on electroweak symmetry breaking. We thus treat top partners in more detail while choosing an effective parametrization for the other quark partners.  The simplest top partner implementation is to introduce one vector-like top partner multiplet $\Psi$ in the $\bf{5}$ of $SO(5)_R$ in the composite (second) site. $\Psi$  mixes with the elementary doublet $q_L=(t_L,b_L)$ and the singlet $t_R$.  The $q_L$ and $t_R$ elementary fermions are embedded in $Q_L$ and $T_R$ as incomplete $SO(5)_L$ fiveplets in the first site. In terms of   $SO(4)$, $\Psi$ can be decomposed into one $SO(4)$ singlet, $\Psi_1=\tilde{T}$ and a fourplet, $\Psi_4$ which itself contains two $SU(2)_L$ doublets $(T,B)$ and $(X_{5/3}, X_{2/3})$. As $SO(5)$ is explicitly broken,  $\Psi_1$ and $\Psi_4$ have different Dirac-fermion masses i.e. $M_1\neq M_4$. 

The elementary top $t_{L,R}$ and the top partners  $(T,X_{2/3},\tilde{T})$ have the same charge, and they mix due to the spontaneous symmetry breaking. The left- and right-handed Standard Model tops are thus ``partially composite'', i.e. they are linear combinations of the elementary top and its partners. We denote the three other heavy  mass eigenstates (``top-partners'') with $T_{f1},T_{f2}, T_s$. For our phenomenological study, our main focus is on the lightest  top partner mass eigenstate $T_{f1}$. In the benchmark points, we choose $M_4 \ll M_1$, and in this case $T_{f1}$ is dominantly  $X_{2/3}$, with sub-leading contributions from $t$ and $\tilde{T}$. 

Partners of quarks other than the top can be introduced in a similar matter, however, their mixing is simpler to parameterize, because non-negligible mixing between quark and quark partners can only present in either the left- or the right-handed sector.\footnote{The product of the left-and right-handed mixing angle is proportional to the respective quark Yukawa coupling, and thus sizable mixings between both the left- and the right-handed quarks only occur for the top.} Also, non-universal mixing among the light quarks leads to tree-level flavor changing neutral currents. We thus follow a simplified parametrization for the mixing of first and second family quarks used in Ref.\cite{Low:2015uha} with a common mixing angle $s_{L,q}\equiv \sin(\theta_L^q)$. 



The full Lagrangian for gauge and matter sector can be found in Appendix~\ref{app:1}. There, we also give the full top-mass matrix and analytic results for its diagonalization to leading order in $v/f$ and further details on the calculation of the couplings in the mass eigenbasis.\footnote{For our benchmark points and our simulation, we do not rely on the leading order results in $v/f$ but instead numerically diagonalize the mass matrices.}

\subsection{Summary of the interactions}

In this article, we focus on the production of the neutral heavy vector $\rho_0$ and its subsequent decay into $\bar{t}T\rightarrow \bar{t} t Z$ (or the conjugate process). The expressions for interactions of the $\rho_0$ with SM particles and top partners are given in Appendix~\ref{app:1}. Here, we just highlight the main  couplings relevant for the process we study, which are
\begin{eqnarray}
g^L_{\rho_0 q\bar{q}} &=& -\frac{\hat{g}^2 }{g_{\rho }}\left(1-\frac{g_{\rho }^2 ~s_{L,q}^2}{\hat{g}^2}\right) \, ,\label{eq:rhoqq} \\
g^{L,R}_{\rho_0 T_{f,1} T_{f,1}} &=&  \frac{3 g_{\rho} c_y -4 \hat{g}_y s_y }{6}\, \label{eq:rhoTT}\\
g^R_{\rho_0 T_{f,1}t} &=& c_y~ s_{R,t}~ \frac{v}{f}~  \frac{g_{\rho}}{2 \sqrt{2}}  \frac{ M_1}{M_4} \,  \label{eq:rhotT} \\
g^R_{T_{f,1}tZ} &=&  s_y~ s_{R,t} ~\frac{v}{f}~~ \frac{g_{\rho}}{2 \sqrt{2}}  \frac{ M_1}{M_4} \, \label{eq:rhotTZ} 
\label{eq:tTZ}
\end{eqnarray}
where $L,R$ denote the chirality of the quarks in the vertex, $\hat{g}$ denotes the  gauge coupling associated with the elementary $SU(2)_L$ field in the first site, $s_{L,q}$ is the mixing angle in the light quark sector, $s_{R,t}$ denotes the dominating mixing angle in the top sector (c.f. Eq.~(\ref{eq:tmix})),\footnote{Note that in the large $M_1$ limit, $s_{R,t}\rightarrow y_R f / M_1$, such that couplings in Eq.~(\ref{eq:rhotT},\ref{eq:tTZ}) stay finite even in this limit.}  $M_1$ and $M_4$ are the Dirac mass parameters of the singlet and fourplet top partner sector,  and finally $c_y$ and  $s_y$ are the cosine and sine of the mixing amongst neutral elementary and composite gauge bosons (c.f.  Eq.~(\ref{eq:gmix})). Eqs.(\ref{eq:rhoqq} - \ref{eq:rhotTZ}) describe the interactions with the lightest top partner, $T_{f,1}$ for our choice of sample points. For the interactions of the other top partners $T_{f,2}, T_s$ {\it c.f.} Appendix~\ref{app:1}. 

The $\rho_0$ of this model is dominantly produced from a $q\bar{q}$ initial state at the LHC through the interaction in the Eq.~(\ref{eq:rhoqq}).\footnote{$\rho_0$ could also be produced via vector boson or vector boson Higgs fusion, but these production cross sections are very small  for the production of a vector resonance with a TeV scale mass.} The coupling $g^L_{\rho_0 q\bar{q}}$ obtains contributions from two sources. The first contribution arises from $q\bar{q}$ production of an elementary gauge boson which mixes with the composite gauge boson. This coupling is a universal coupling of the $\rho_0$ to all SM fermions (including leptons) which couple to the $Z$ boson. The second contribution can be thought of as a strong-sector coupling of the $\rho_0$ to composite quarks, which mix with the elementary quarks through the mixing angle $s_{L,q}$. 

For the decay of the $\rho_0$, it is possible to have a scenario where  $\rho_0$ decays to top and its partner dominate over top partner pairs. This occurs for  two reasons: first, the  coupling $g^R_{\rho_0 t T_{f,1}}$ can be larger than $g^R_{\rho_0 T_{f,1} T_{f,1}}$, even though the interaction of $\rho_0$ to top partners is of  O($g_\rho$). This is because the coupling of $\rho_0$ to a top and a top partner also originates from the strong sector and is enhanced in the limit when the singlet top partner is decoupled from the fourplet  and right handed top is mostly composite (making the mixing angle $\mathcal{O}(1)$). Second, even when the top partner pair decay is kinematically open, the branching ratio stays smaller because of the suppression by the phase space factor. 




Couplings of the $\rho_0$ to other Standard model fermions  are either of electroweak strength (as opposed to strong-coupling),  suppressed by an additional gauge boson mixing angle, or (in the case of light quarks), by $s^2_{L,q}$.  Thus, if a decay involving top partners is allowed, two-body decays of the $\rho_0$ into Standard Model particles are generically suppressed.



	 
 \subsection{Collider Constraints}\label{subsection:constraints}
 In this section, we summarize bounds on parameters and mass scales relevant for the phenomenological studies of interest. For earlier discussions of bounds {c.f.} also Refs.~\cite{Panico:2011pw,Greco:2014aza,Low:2015uha}.
 
\subsubsection{Indirect Constraints on Composite models}\label{sec:13tevindirect}

\paragraph{Higgs couplings} ~\newline
Composite models generally predict deviations of the Higgs couplings to the SM particles. These deviations depend on the choice of coset group and embeddings of fermions. The Higgs couplings to SM particles (massive gauge bosons $V \equiv Z,W$ and fermions) and Higgs self couplings are tabulated in Refs.~\cite{Carena:2014ria,Grober:2016wmf}.  The  current  constraints of Higgs coupling deviations from the Standard Model value from a combined ATLAS and CMS analysis of the LHC pp collision data at $\sqrt{s}$ = 7 and 8 TeV~\cite{Khachatryan:2016vau}  are $\lesssim 10 \% $ . 
Using Higgs couplings in Minimal Composite Higgs Model (MCHM5) to gauge couplings normalized to the  corresponding SM couplings, i.e $g_X /g_X^{SM} = \sqrt{1-\xi}$, where $\xi=(v/f)^2$ puts $f$ to be above 550 GeV.  However we take the conservative bound on $f$ from the projections of the High Luminosity LHC Run, i.e. $f > 800$ GeV~\cite{Thamm:2015zwa}. 
\titlespacing*{\subparagraph}
 {0pt}{0.3\baselineskip}{3\baselineskip}
 
\paragraph{Electroweak Precision Tests (EWPT)} 

\subparagraph{\textit{Oblique parameters}} ~\newline
Composite Higgs models usually contain extra contributions to $\hat{S}$ and $\hat{T}$  electroweak precision variables~\cite{Grojean:2006nn,Pich:2012dv,Azatov:2013ura,Bunk:2013uea,Pich:2013fea,Grojean:2013qca,Bellazzini:2014yua}. Constraints on $\hat{S}$ can be used to provide lower bounds on the mass of composite vector resonances. The $\hat{S}$ parameter receives tree level corrections from the presence of the composite vector states, sub-leading ``universal" 1-loop corrections, which are fixed by the infra-red (IR) dynamics~\cite{Barbieri:2007bh}. In addition, logarithmically divergent  one-loop corrections from the fermions also appear, but  are generally model dependent~\cite{Grojean:2013qca,Matsedonskyi:2012ym}. In the two site construction, it is possible to estimate the leading fermionic contributions, but it becomes necessary to include the dependence on mixing angles on the top sector  ~\cite{Panico:2015jxa}, making it difficult to put definite limits on $m_{\rho}$ and $f$. Finally, there can be large finite contributions to the $\hat{S}$ parameter from the UV dynamics of the theory.  

The $\hat{T}$ parameter obtains no tree level contributions from the vector resonances owing to custodial symmetry.  It does get sizable positive contributions from fermion loops (which depend on fermionic embedding).  Additional sub-leading negative contributions to $\hat{T}$ parameter come from vector resonances.  Thus even though  EWPT provides indirect bounds on $m_{\rho}$ and decay constant $f$ ~\cite{Panico:2015jxa}, the bounds are parameter dependent and have a sizable uncertainty due to the UV dependence\footnote{Note that current estimates of the EWPT variables still allow for $m_{\rho}\sim 2$ TeV~\cite{Barducci:2015vyf}.}. We therefore use bounds on $f$ and $m_\rho$ following from bounds on modifications on Higgs couplings, and direct searches for vector resonances instead.


\subparagraph{\textit{Modified Z Couplings on bottom and light quarks}} ~\newline
Oblique parameters encode universal NP effects which affect all fermion generations in the same way.  As strong dynamics plays a crucial role in generating a large top mass, the observables related to the third quark generation receive non-universal corrections. Corrections to the coupling of  $Z$ to  $b$ quarks are well-constrained. However, given that the two site model mostly respects left-right symmetry, the $Zb\bar{b}$ is protected and bounds from electroweak precision variable, $R_b ~(\Gamma_{Z\to b\bar{b}}/ \Gamma_{Z\to had})$ are relaxed~\cite{Panico:2015jxa,Low:2015uha}. 

For the light quarks, as already mentioned,  we assume partial compositeness and set the left handed mixing of the lightest two generations of quark doublets to a common value of $s_{L,q}$.  Light quarks can have sizable mixing due to the flavor symmetry which relates their mixing to the mixings of the third generation quarks. This mixing leads to a modification of the couplings of the light generations which is constrained from the measurements of total hadronic Higgs width, $R_h$~\cite{Redi:2011zi}. Furthermore, unitarity of the CKM matrix  puts significant bounds on left-handed couplings of the up sector ~\cite{Redi:2011zi}. At leading order the bounds from $R_h$ and CKM unitarity (specifcally, $\delta V_{ud}$~\cite{Matsedonskyi:2014iha}) are correlated and proportional to $\delta g_{uL}$ and allows for $s_{L,u} < 0.15$. The bound from quark compositness is much milder~\cite{Low:2015uha}.

\paragraph{Flavour Constraints on Top sector}~\newline
We have kept the lagrangian of  top partners  in the $\mathbf{5}_L+\mathbf{5}_R$ model.  We focus on flavour symmetric scenario, with  $U(2)^2$ horizontal symmetry~\cite{Low:2015uha,Redi:2011zi,Barbieri:2012tu,Matsedonskyi:2014iha} for the first two families, in the left handed mixings. Dominant contributions to tree level $\delta F=2$ processes from four quark interactions also exist. The most relevant four fermion operators involve down-type quarks and have a form~\cite{Low:2015uha,Matsedonskyi:2014iha}
\begin{equation}
C_{d_L} (V_{3i}^* V_{3j})^2 C_{ij}^2 \frac{s^4_{L,t}}{f^2} (\bar{d^i_L}\gamma^{\mu}d^j_L)(\bar{d^i_L}\gamma_{\mu}d^j_L)
\end{equation}
where  the coefficient $C_{dL}$ is expected to be order one~\cite{Barbieri:2012tu}, $V_{ij}$ are the elements of CKM matrix and lastly,  the coefficients $C_{ij}$ depend on the flavor pattern.  The emerging bound on $s_{L,t}$ has a dependence of $(C_{d_L})^{1/4}$ and is thus very insensitive to uncertainties on this coefficient's size.  For $U(2)^2$ flavor symmetry in left handed couplings, $\Delta B_s =2$  observables are  most constraining, and so we focus on $C_{23}$, which is a free parameter for $U(2)^2$ scenario. For $m_{\rho} \sim 2$ TeV , the bound on top sector's left handed mixing is
\begin{equation}
s_{L,t} \lesssim 0.95 \left(\frac{3}{g_{\rho}}\right)^{1/2}\, .
\label{eq:sLtbound}
\end{equation}
All the benchmark values of  $s_{L,t}$  we consider satisfy the above bound. 

\paragraph{Lepton Sector} ~\newline
We treat the leptons as elementary , and consider them massless as well. All the couplings of leptons to spin-1 particles originate universally from the vector mixing. There are no additional indirect constraints coming from the lepton sector.

\subsubsection{Direct Experimental constraints on $\rho_0$ decay channels}
We ensure that our benchmark points are not excluded by ATLAS and CMS searches for its decay channels other than single top partner production associated with one top quark. Decay channels constrained from resonance searches are dileptons,  \cite{ATLAS:2016cyf,CMS:2016abv}, top~\cite{ATLAS:ttbar,CMS:2016zte} and bottom \cite{ATLAS:2016gvq} pair production, di-jets \cite{ATLAS:2016lvi,CMS:2016wpz}, and di-bosons \cite{ATLAS:2016cwq,ATLAS:2016kjy,ATLAS:2016yqq,CMS:2015nmz,CMS:2016pfl}.  The bounds on production cross section times branching ratio in the resonant $jj$, $bb$ and $t\bar{t}$ channel in the $m_\rho \sim 2$ TeV regime are above 100 fb and we ensure that our benchmark points do not exceed this bound.

Currently, the most stringent upper limits (at 95 \% CL) for a 2 TeV resonance from the LHC run 2 at $\sqrt{s}=$ 13 TeV and  13 fb$^{-1}$ come from dileptons, $\sigma~\times~\br~(pp~\to~Z'~\to~\ell^+~\ell^-)<0.34\fb$ \cite{ATLAS:2016cyf} and for $\sigma \times \br (pp \to Z' \to WW) < 4.16$~fb \cite{ATLAS:2016cwq}.

 
 \subsubsection{Direct Experimental constraints on $T_{f1}$ decay channels}
 
 As the composite mass scales $M_4$ and $M_1$ arise from the condensation of the strong sector, they need to lie between $f$ and $4 \pi f$.  Searches for top partners with exotic charge $5/3$ at 13 TeV,  in channels with same-sign di-leptons  and  lepton+jets, have placed lower limits on mass of $X_{5/3}$ at 960 GeV~\cite{CMS:2015alb}.  This requires that we consider the fourplet top sector mass term $M_4 >960$ GeV (as $m_{X_{5/3}} = M_4$). Top partner searches at 8~TeV have already excluded heavy colored fermions in the 500-950 GeV window, depending on the quantum numbers and on the branching ratios~\cite{Khachatryan:2015oba,Aad:2015gdg,Aad:2014efa,Aad:2015kqa,CMS-PAS-B2G-16-001}, so it is safe to keep a lower bound of 1 TeV on top partners. Fixing the top mass to be close to its observed value needs $y_L$ and $y_R$ to be $\mathcal{O}(1)$. \newline

\subsection{Benchmark Points} \label{sec:benchmark}

To fix the benchmark points, we seek parameter points which are not excluded by (but are close to the current sensitivity of) the current experimental bounds. We choose  
\begin{equation}
 g_{\rho}= 3.5 ,~~ f=808 ~\text{GeV},~~ m_{\rho}= 2035 ~\text{GeV} ,~~ M_1= 20 ~\text{TeV}, ~~s_{L,q}= 0.1
 \label{eq:BMfix}
 \end{equation}
and three different values of $M_4$ (which to leading order is the mass of the top partner $T_{f1}$) and $y_r$, given in Table~\ref{tab:BMdiff}.\footnote{With these parameters fixed, the left-handed pre-Yukawa coupling $y_L$ is determined by the requirement that the top has the correct mass. We checked that for all parameter points that $y_L < 4 \pi$, i.e. that the coupling is not excessively large and non-perturbative.} 

\begin{table}[h!]
	\begin{center}
		\begin{tabular}{ | c | c | c | c | }
			\hline
			 &  SP1 & SP2 & SP3  \\ \hline
			$M_4$ [GeV]& 1000 & 970 & 1030   \\ 
	        $y_R$ & 10 & 11 & 11 \\
			\hline
			\end{tabular}
			\caption{Remaining parameters for the benchmark points.}
			\label{tab:BMdiff}
			\end{center}
			\end{table}

The choice of $f$ satisfies the bound $f \gtrsim 800$ GeV. The choice of $g_\rho = 3.5$ then yields a mass of $m_{\rho}= 2035 ~\text{GeV}$, i.e. around 2 TeV,  through Eq.~(\ref{eq:spectrum}). We will comment below how the bounds from $Z'$ searches are satisfied for this resonance mass. The choice $s_{L,q}= 0.1$ is in accord with bounds on light-quark compositeness. Finally, the choice $M_1 = 20$~TeV is taken (mostly) as a simplification -- as we choose $M_1\gg M_4$, only two of the three top partners $T_{f1}, T_{f2}, T_s$ are in the TeV regime, while the third partner is much heavier and nearly decoupled.  

The choices for $M_4$ and  $y_R$ are taken to illustrate cases in which $\rho_0$  decay into top partner pairs is open, ($m_{T_{f1}}< m_{\rho}/2$) or not, as this can a priory make a large difference. The detailed of the mass spectrum, and the parameter fixing of $y_L$ in order to reproduce the correct top mass are given in Appendix \ref{app:2}.\footnote{ {\it C.f.} Table~\ref{tab:masses2} for $y_L$ and the mass spectrum as well as Tables~\ref{tab:couplings2} - \ref{tab:BRTf12} for couplings and relevant  production rates and branching ratios.} In the case of the first benchmark point, SP1, single top partner production in association with the top is maximal.
In SP2, the top partner pair production is kinematically allowed but is small because of phase space suppression.  For the last benchmark point, SP3, top partner pair production is closed, but the single top partner associated with top has a lower branching ratio than SP1.  
Note that while our benchmark points contain a partially composite implementation for the $t_R$, our choice of large $y_R$ should allow us to relate to scenarios where right-handed top is fully composite~\cite{Matsedonskyi:2015dns,Delaunay:2013pwa}. However a detailed analysis of the structure of the couplings in a partial versus fully composite $t_R$ is beyond the scope of this study. 

\begin{figure}[h]
	\begin{center}
		\includegraphics[width=0.6\textwidth]{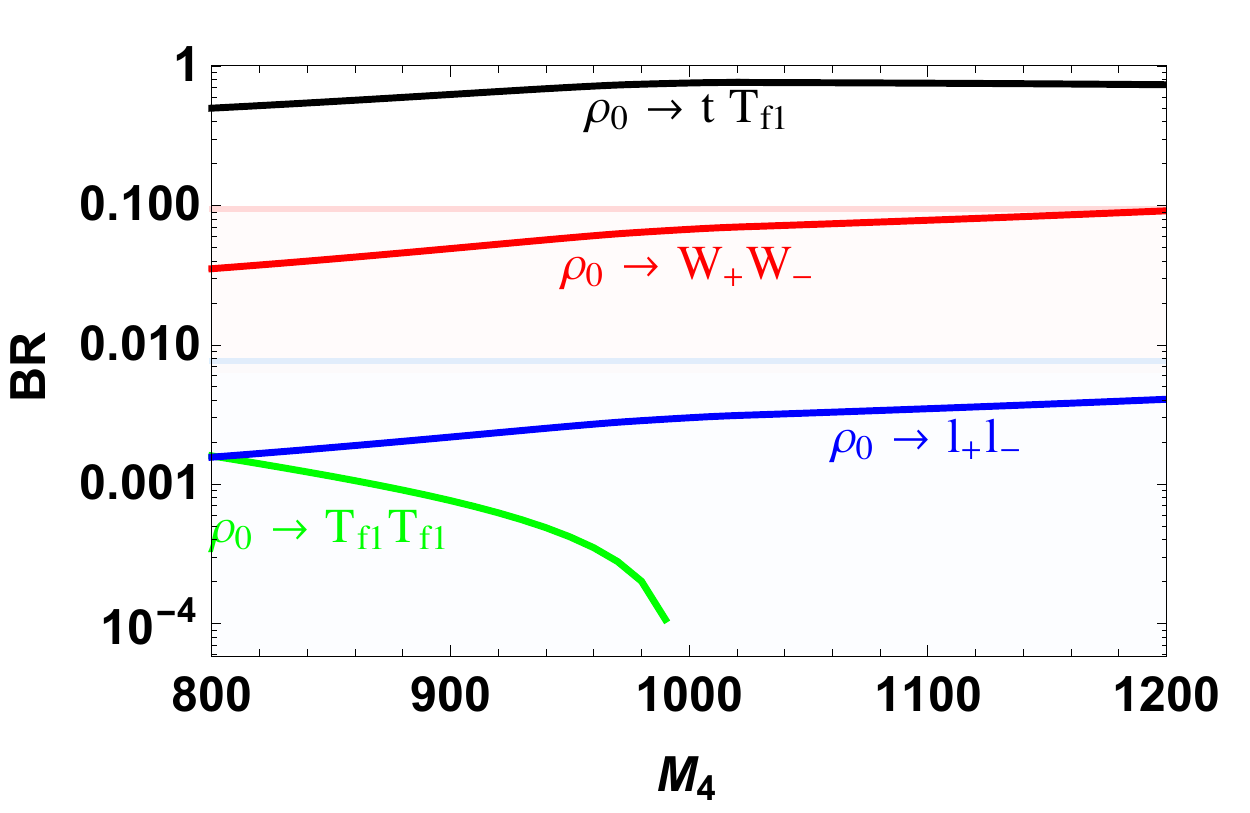}
		\caption{The relevant heavy vector resonance branching ratios as function of $M_4 \sim m_{T_{f1}}$. Model parameters other than $M_4$ are fixed to the parameter values given in Eq.~(\ref{eq:BMfix}), and $y_R = 10$. The various curves correspond to the  following decay channels:  $t\bar{T_{f1}}+\bar{t}T_{f1}$(black); $W_{+}W_{-}$ (red); $l_{+}l_{-}$ (blue); and lastly  $T_{f1}T_{f1}$ .The light red and blue filled regions indicate the current constraints from diboson~\cite{ATLAS:2016cwq} and dilepton~\cite{ATLAS:2016cyf}  searches. }
		\label{fig:BRfun_M4}
	\end{center}
	
\end{figure}

Figure.~\ref{fig:BRfun_M4} shows several branching ratios  for the decay of  a 2 TeV $\rho_0$ with the parameter choices given in Eq.~(\ref{eq:BMfix}) and with $y_R = 10$ as a function of $M_4$. Single top partner production is dominant, while the di-boson and (in particular) di-lepton branching ratios are (severely) suppressed. For our benchmark points we find production cross sections $ \sigma(p p \rightarrow \rho_0) \approx 52 $ fb. Together with branching ratios from Fig.~\ref{fig:BRfun_M4} we see that the di-boson \cite{ATLAS:2016cwq} and di-lepton bounds \cite{ATLAS:2016cyf} from $Z'$ searches are satisfied, while the production of a top and its partner through the $\rho_0$ resonance has a production cross section of $\sim 38$ fb.

\begin{figure}[h]
	\begin{center}
		\includegraphics[width=0.6\textwidth]{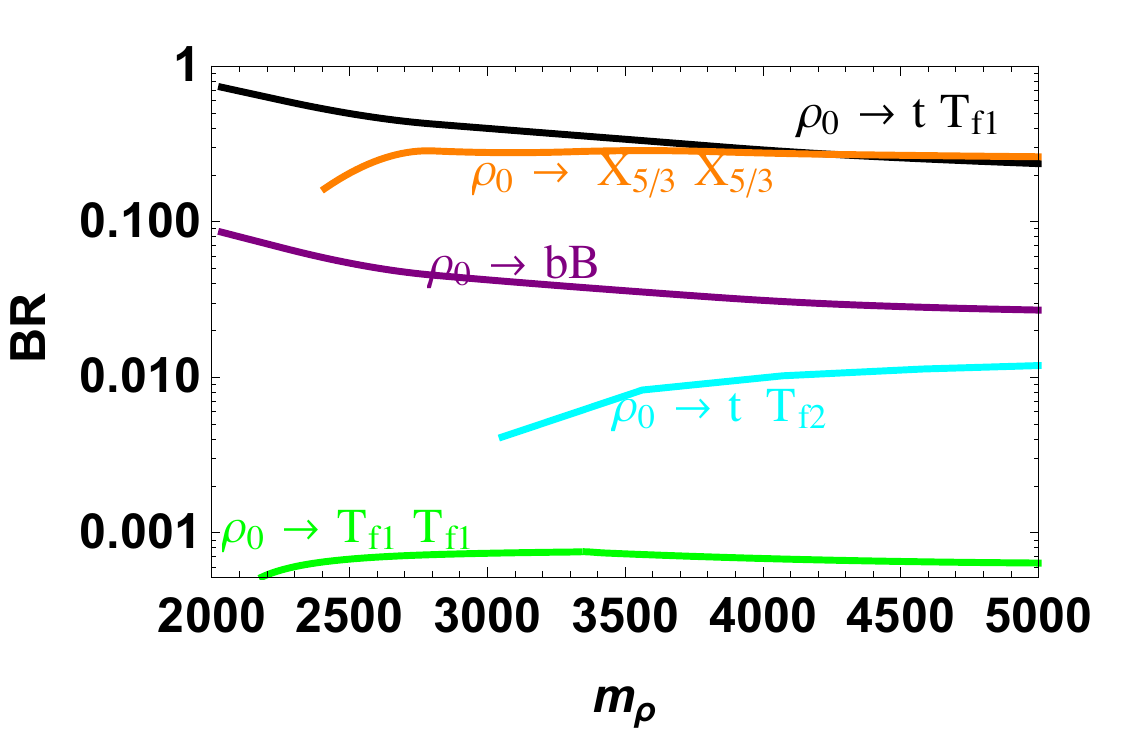}
		\caption{Branching ratios of vector resonance as function of $m_ {\rho}$. We vary $m_{\rho}$ by changing f, while keeping $y_R = 10$, $M_1=20$ TeV, $M_4= 970$ GeV,  $s_{L,q}= 0.11$ and $g_{\rho}=3.5$.		
		The various curves correspond to the  following decay channels: $t\bar{T_{f1}}+\bar{t}T_{f1}$(black); $X_{5/3}X_{5/3}$ (orange);  $b\bar{B}+\bar{b}B$ (purple) ;$t\bar{T_{f2}}+\bar{t}T_{f2}$ (cyan); and lastly, $T_{f1} T_{f1}$ (green).}
		\label{fig:BR_funmrho}
	\end{center}
\end{figure}

As we study in detail a collider search for the neutral vector resonance, it is useful to emphasize that the decay channel with $t\bar{T_{f1}}+\bar{t}T_{f1}$ \textit{dominates} over the other remaining top partner decay modes even beyond the $m_{\rho} = 2 m_{T_{f1}}$ threshold. This occurs even in the high $m_{\rho}$ region where there is no phase space suppression in the $T_{f1} T_{f1}$ decay mode. This can be understood from the choice of large values of $y_R$ and $M_1$, which yield a larger right handed coupling of  $\rho_0$ to $t T_{f1}$ as compared to its counterpart, $T_{f1}T_{f1}$. We show as an example, the branching ratios for the decay of $\rho_0$ in Figure.~\ref{fig:BR_funmrho} as a function of the mass of $m_{\rho}$ when we vary $m_{\rho}$ by changing f, while keeping $y_R = 10$, $M_1=20$ TeV, $M_4= 970$ GeV,  $s_{L,q}= 0.11$ and $g_{\rho}=3.5$. The states $X_{5/3}$ and $B$ for which we show BR in Figure.~\ref{fig:BR_funmrho}, are top partners with charge $5/3$ and $-1/3$ (c.f. Appendix~\ref{app:1} for details on the top partner embedding).

According to its quantum numbers, the top partner $T_{f1}$ can decay further into $bW$, $tZ$ or $th$. The branching ratios for charge $2/3$ partners are strongly parameter dependent. This was discussed in some detail in Appendix B. of Ref.~\cite{Backovic:2014uma}, where it was shown that for charge $2/3$ top partners in the $\bf{5}$ , branching ratios for the decays of $T_{f1}$ into $tZ$ and $th$ scale as  $\sim M_1^2/(M_1^2+M_4^2)$ and  $\sim M_4^2/(M_1^2+M_4^2)$ respectively. Thus our choice of $M_1\gg M_4$ yields dominant decay $T_{f1}\rightarrow tZ$ and a dominant final state $\rho_0 \rightarrow \bar{t}tZ$ which we discuss in detail in the next section. For $M_1\gtrsim M_4$, the $T_{f1}\rightarrow tZ$ branching ratio would be reduced at the expense of  $T_{f1}\rightarrow th$ which would yield an (also very interesting) final state $\rho_0\rightarrow t\bar{t}h$.  However, we  leave the  detailed studies of this final state for the future.

\section{Collider Phenomenology}\label{sec:pheno}

  In this section we study the ability of the LHC13 run to probe the  scenarios of boosted $t\bar{t}Z$ event topologies. Scenarios in which the top partner is produced via decays of a heavy vector resonance result in interesting, previously overlooked, LHC signatures. As the vector resonance undergoes a three body decay into $t\bar{t}Z$, and the mass of the resonance is $\mathcal{O}$(few) TeV, the final state of interest will be characterized by a highly boosted ``triplet'' of heavy SM states. The boosted event topology is relatively straightforward to reconstruct, even though it consists of at least 8 particles in the final state (6 for the decays of two tops and 2 from the decay of the $Z$). The large boost of the heavy SM particles results in their decay products being kinematically clustered together, hence reducing any possible combinatorical issues in reconstructing the event.  For the purpose of illustration, we will consider several benchmark points of the two site composite Higgs model which are not excluded by the existing collider results. Note however that our collider analysis can easily be extended to any search in the boosted $t\bar{t}Z$ channel.
  
We rely on the {\sc MG5\_aMC} \cite{Alwall:2014hca} framework for event generation at parton level, while we shower the events using {\sc Pythia 6} \cite{Sjostrand:2006za}. 
We generate all signal events at leading order, while for SM backgrounds we use samples matched to one extra jet. In order to increase the statistics in the background event samples, we impose a cut of $H_T > 800 \GeV$ on the hard processes level. All event generation assumes the leading order {\sc NN23PDF} set~\cite{Pumplin:2002vw} with the default {\sc MG5\_aMC} normalization/factorization scale.  We cluster the showered events using the {\sc fastjet} \cite{Cacciari:2011ma} implementation of the anti-$k_T$ algorithm, using a  $R=1.5$ jet cone for the purpose of the ``fat jet'' analysis, while for the purpose of $b$-tagging, we employ the standard cone of $r=0.4$. \footnote{Note that reducing the jet cone in the fat jet (fj) analysis to, say, $R=1.0$ could help in mitigating pileup and even reduce the overall background contributions. Here we chose $R=1.5$ in order to be in line with the CMS top tagging recommendations of Ref.~\cite{CMS:2014fya}.} Our simulations do not consider detector effects or pileup. 

Throughout the paper we employ simplified $b, Z$ and top tagging procedures\footnote{See Refs.~\cite{Butterworth:2008iy, Almeida:2011aa, Backovic:2012jj, Schlaffer:2014osa, Ellis:2007ib, Abdesselam:2010pt, Salam:2009jx, Nath:2010zj, Almeida:2011ud, Plehn:2011tg, Altheimer:2012mn, Soper:2011cr, Soper:2012pb, Jankowiak:2011qa, Krohn:2009th, Ellis:2009me, Backovic:2012jk, Backovic:2013bga, Hook:2011cq, Thaler:2010tr, Thaler:2011gf, Thaler:2008ju, Almeida:2008yp,Almeida:2010pa, Rentala:2014bxa, Cogan:2014oua, Larkoski:2014wba,Almeida:2015jua,Moult:2016cvt,Baldi:2016fql,deOliveira:2015xxd,Larkoski:2014zma} for details of other tagging methods used to classify and distinguish boosted heavy objects such as Higgs, top and W/Z bosons from the QCD backgrounds}. For the purpose of $b$-tagging, a $r=0.4$ jet in the event is tagged as a $b$-jet, based on the presence of a parton level $b$ quark within $\Delta R < 0.4$ from the jet axis, weighted by the appropriate $b$-tagging efficiency.  For the purpose of our numerical analysis, we consider the $b$-tagging benchmark of 
\begin{equation}
	\epsilon_b = 0.70\,, \,\,\,\,\, \epsilon_c = 0.18\, , \,\,\,\,\, \epsilon_j = 0.017\, , \nonumber
\end{equation}
where $\epsilon_{b, c, j}$ are the probabilities that a $b, c$ or light jet will be tagged as a $b$-jet. 

For the purpose of boosted top tagging use the CMS benchmark point ~\cite{CMS-PAS-JME-15-002} of 
$$
	\eps_t = 0.5 \,, \,\,\,\,\, \epsilon_j = 0.005\,, 
$$
while for boosted $Z$ tagging we use the CMS benchmark \cite{CMS:2014joa} of
$$
	\eps_Z = 0.5, \, \,\,\,\,\,\epsilon_j = 0.03\, ,
$$
where $\epsilon_{Z, j}$ are the probabilities that a $Z$ boson or a light jet will be tagged as $Z$ boson respectively. Note that the top tagging efficiencies include fat jet $b$-tagging.

\subsection{Final States with 2 Leptons and no Missing Energy}

The simplest event topology consists of final states in which both top quarks decay hadronically while the $Z$ boson decays into a pair of muons or electrons (for a brief discussion of some other final states see Section~\ref{sec:otherchannels}). The leptonic $Z$ final state suffers from a smaller  BR$(Z\rightarrow l^+l^-) \approx 7\,\%$ compared to the fully hadronic channel, but also comes with several important advantages. First, the lack of missing energy in the event allows for full reconstruction of the event, as well as the vector resonance and the top partner masses. Second, the only relevant background channel for event topologies with two boosted, hadronically decaying tops and a $Z$ boson is the SM $Z$+jets production. Other SM backgrounds like $t\bar{t}$ contribute about 10 \% of the total background, while SM $t\bar{t}Z$ is negligible at high event $H_T$. The before-mentioned features make the leptonic $Z$ channel with two hadronically decaying boosted tops a very attractive channel for new physics searches. 

\begin{figure}[h]
\begin{center}
 \includegraphics[width=0.41\textwidth]{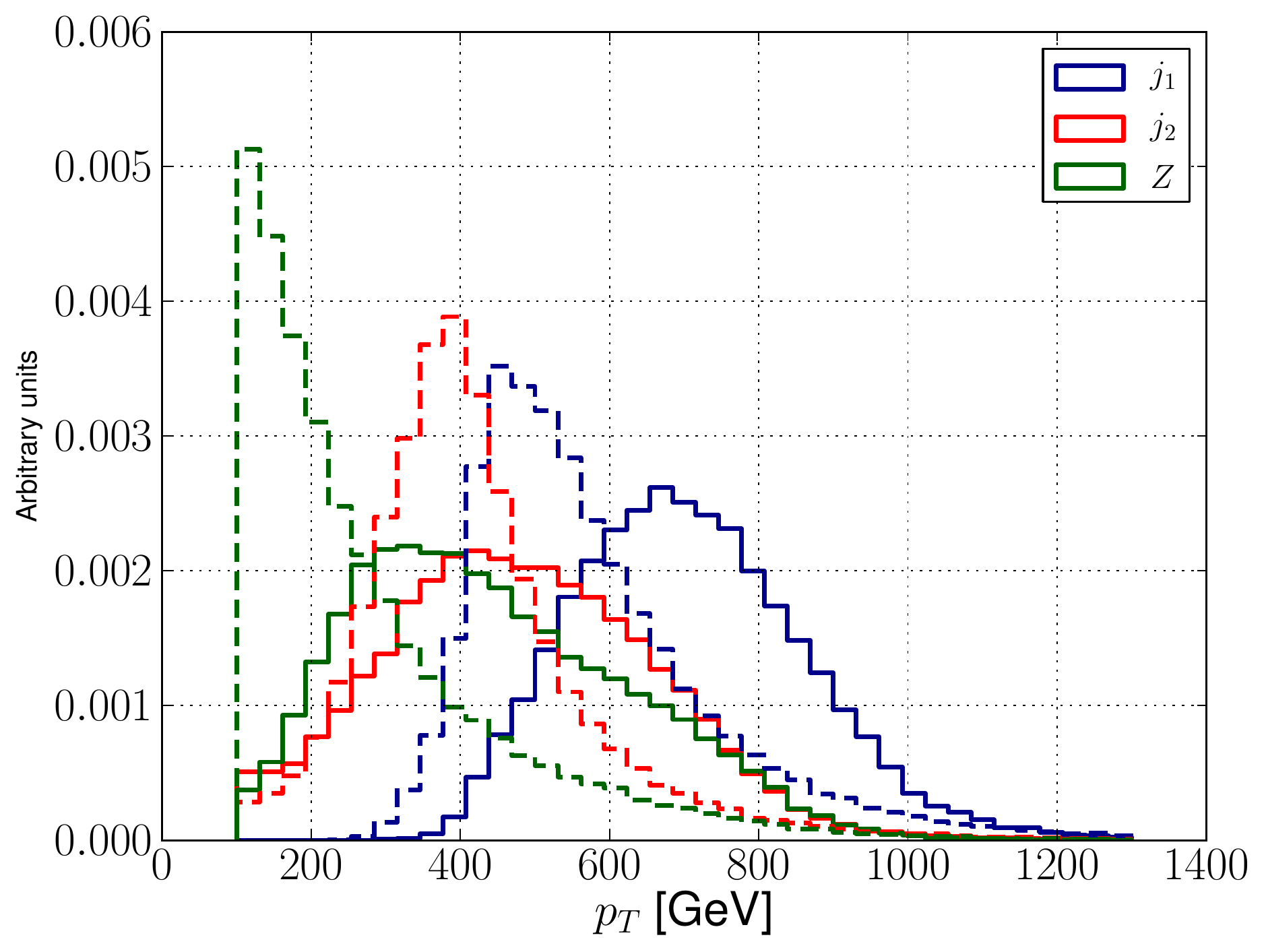}
 \includegraphics[width=0.4\textwidth]{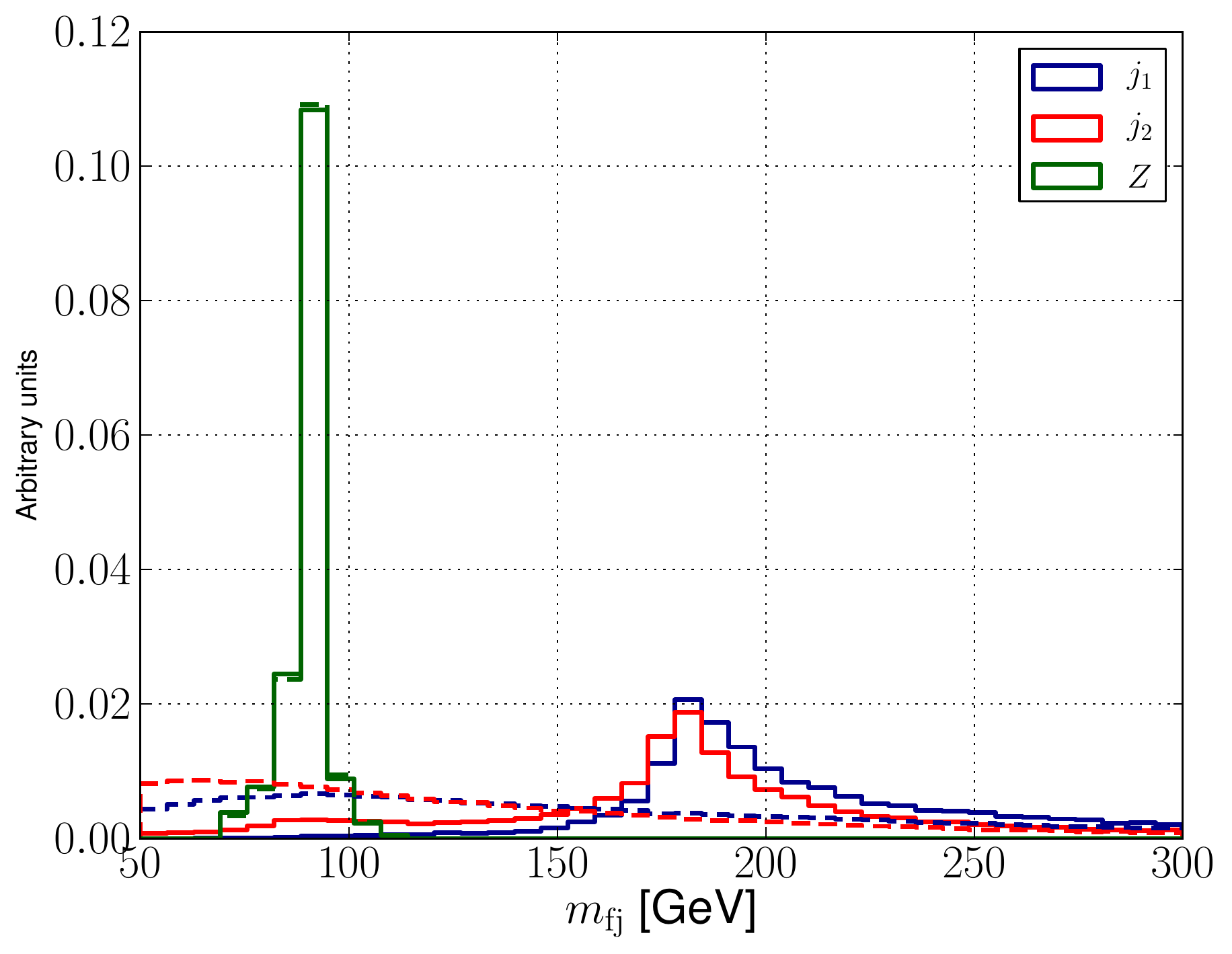}
 \includegraphics[width=0.4\textwidth]{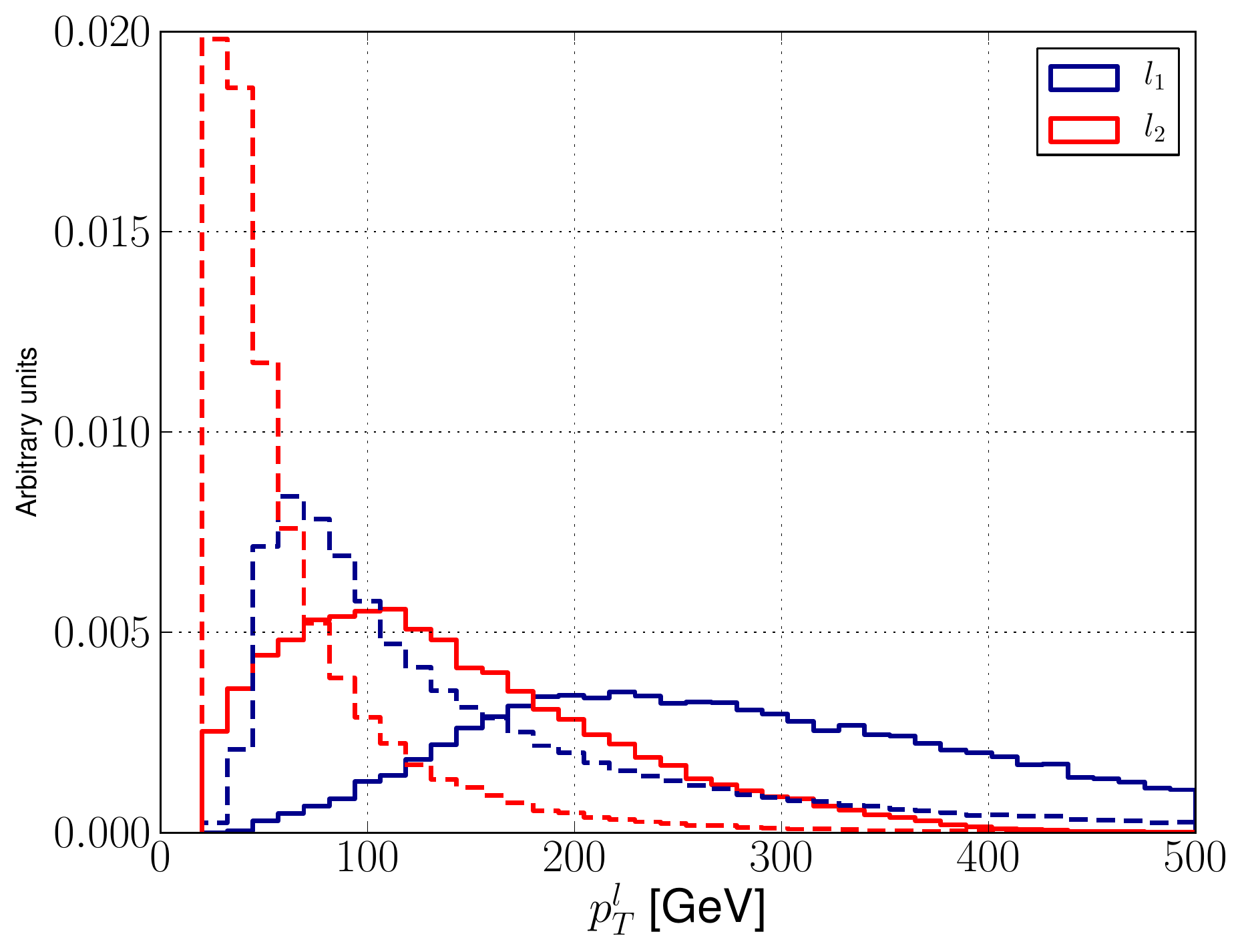}
 \includegraphics[width=0.41\textwidth]{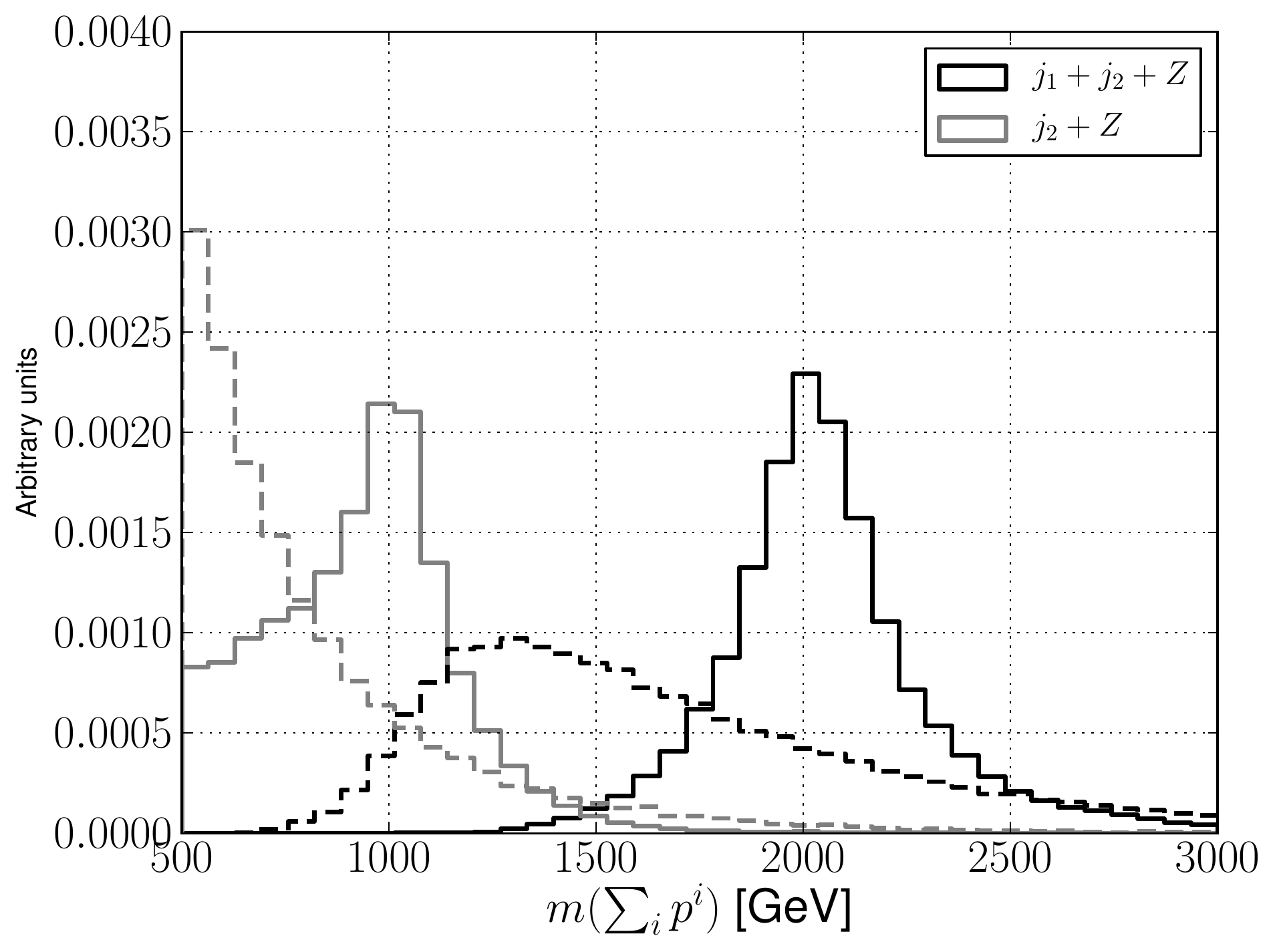}
\caption{Distributions of kinematic observables for signal (solid lines) and background (dashed lines) events. For the purpose of illustration, we show the events generated for the SP1 model point while the background mostly consists of $Z$+jets events. Labels $j_{1,2}$ refer to the hardest and second hardest $R=1.5$ jet in the event, and $Z$ represents the sum of two hardest leptons in the event ($l_{1,2}$). The events assume no pileup, detector simulation or top tagging. All distributions are normalized to unit area.}
\end{center}
\label{fig:kinematics}
\end{figure}

 Figure~\ref{fig:kinematics} shows relevant kinematic distributions of the signal and background events, where dashed lines correspond to background events and solid lines represent the SP1 model point for the purpose of illustration. The top left panel shows the transverse momentum distributions for reconstructed jets and the $Z$ boson. Signal events show a significantly harder spectrum for the hardest and second hardest jet.  The reconstructed background $Z$ boson's transverse momentum distribution sharply falls off with the increase in $p_T$, while the signal events are characterized by $Z$ bosons of higher average $p_T$. Invariant masses of reconstructed objects, shown in the top right panel of Fig.~\ref{fig:kinematics}, display characteristic peaks at masses of the $Z$ boson and top quarks in case of signal events, while the fat jet invariant mass resembles a continuum distribution in the case of background events. Naturally, the reconstructed $Z$ mass in case of the backgrounds is indistinguishable from the signal events as a true $Z$ boson is present in the background events.  Signal events also show a significantly harder spectrum of both the hardest and the second hardest leptons, as shown in the bottom left panel of Fig.~\ref{fig:kinematics}. Finally, in case of signal events, we expect the invariant mass of the $Z$ and the second hardest jet ($m_{23}$)to reproduce the mass of the top partner, while the invariant mass of the hardest, second hardest jet and the $Z$ ($m_{123}$) should peak at the mass of the vector resonance. Conversely, no resonant structure is expected in the $m_{123}$ and $m_{23}$ distributions of the background events, as seen in the lower right panel of Fig.~\ref{fig:kinematics}. \footnote{Note that the apparent ``peak'' in the invariant mass distribution of the hardest fat jet , second hardest fat jet and the $Z$ boson sum in case of $Z$+jets background is due to the $H_T > 800 \GeV$ cut imposed at generation level in order to improve the background event statistics.}

Next, we study the ability of the LHC to probe the doubly resonant $t\bar{t}Z$ event topologies. Table~\ref{tab:cutflow} shows an example cutflow for three model benchmark points and the $Z$+jets background. The background cross section assumes a conservative $K$-factor of 1.3, while we consider only a leading order cross section for the signal. We begin the event pre-selection by requiring a presence of two $R=1.5$ fat jets in the event with $p^{\rm fj}_T > 100\,\GeV, \, |y^{\rm fj}| < 2.5$, accompanied with two hard leptons with $p_T^l>20 \,\GeV,\, |y^{l}| < 2.5$.  As we expect a $Z$ boson in the final state, we require the di-lepton system to have an invariant mass of $70\, \GeV < m_{ll} < 110\, \GeV$. 

We find that further requiring two high $p_T$ fat jets in the event (in our case with $p_T > 400, 300 \GeV$ respectively) provides good background rejection power, with the factor of $\sim 7$ reduction in background contamination at $\sim 60 \%$ signal efficiency. Top tagging proves to be crucial in bringing the signal to background ratio to be $O(10)$, while defining the signal region with a lower cut on the reconstructed top partner and $\rho_0$ masses can further suppress the SM background at an additional $~ 70 \%$ signal efficiency. Our example cutflow shows that the benchmark points we consider in this paper are discoverable at the LHC13 in the $Z\rightarrow l^+ l^-$ channel with as little as 30 $\fb^{-1}$ of integrated luminosity.

\begin{table}[htb]
\begin{center}

\begin{tabular}{ccccc}

$Z \rightarrow l^+l^-$ & $\sigma$(SP1) & $\sigma$(SP2) & $\sigma$(SP3) & $\sigma$($Z$+jets)   \\
\hline \hline
Preselection & 0.64 & 0.64 & 0.64 & 326 \\
$p_T^{Z} > 300 \GeV$ & 0.48 & 0.46 & 0.49 & 254 \\
$p_T^{j_{1,2}} > 400, 300  \GeV$ & 0.38 & 0.36 & 0.39 & 38 \\
CMS top tag &  0.098 &  0.090 &  0.098 & $9.5 \times 10^{-3}$ \\
\hline
$m_{23} > 800 \GeV$ & 0.074 & 0.074 &  0.074 & $3.5 \times 10^{-3}$   \\
$m_{123} > 1.8 \TeV$ & 0.066 & 0.066 & 0.066 & $2.9 \times 10^{-3}$ \\
\hline \hline
$S/B$& 20 & 20 &  20  &  \\
$S/\sqrt{B} (30  \fb^{-1})$& 6.5 & 6.5 & 6.5 &  \\
$S/\sqrt{B} (100  \fb^{-1})$& 11.8 & 11.8 & 11.8 &  \\
\hline
\end{tabular}
\end{center}
\caption{Example cutflow for the $t\bar{t}Z$ resonance search in the $Z\rightarrow l^+ l^-$ channel, assuming the $t,\bar{t}$ quarks decay hadronically. All samples assume a $H_T > 800 \GeV$ cut at the event generator level. All cross section values are in fb. The background cross section includes an NLO $K$-factor of 1.3. }
\label{tab:cutflow}
\end{table}

\begin{figure}[h]
\begin{center}
 \includegraphics[width=0.7\textwidth]{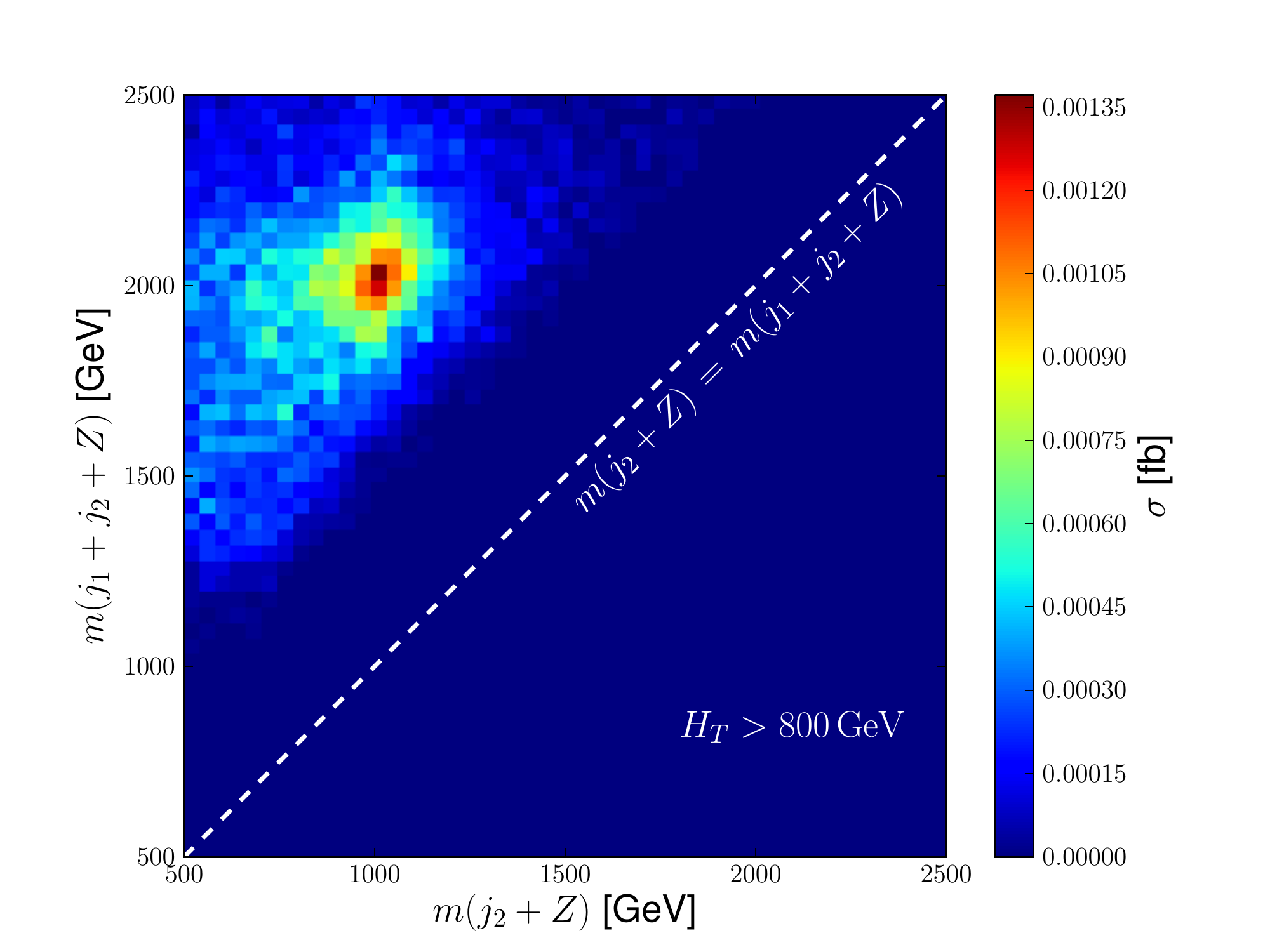}

\caption{Distribution of signal and background in the $m_{23}$, $m_{123}$ plane, where $m_{23}$ refers to the invariant mass of the second hardest top and the $Z$ boson, while $m_{123}$ represents the invariant mass of both tops and the $Z$ boson. The plot shows the scenario of benchmark point SP1 from Table~\ref{tab:cutflow}, with the event selection up to the ``CMS top tag''.}
\end{center}
\label{fig:dalitz}
\end{figure}

The boosted $t\bar{t}Z$ signal topology features a characteristic resonant structure in the $m_{123} \equiv m(j_1 + j_2 +Z)$ and $m_{23}  \equiv m(j_2 +Z)$ plane, which can be particularly useful at identifying the signal events. Figure~\ref{fig:dalitz} shows a Dalitz-like plot of the signal and background normalized to the total $\sigma_S+\sigma_B$. The bulk of the background events lie in the region of $m_{23} \sim 500 -1000 \GeV$ and $m_{123} \sim 1-2 \TeV$, while the signal events show up as a ``blip'' at $m_{23} \sim 1 \TeV$ and $m_{123} \sim 2 \TeV$. The intrinsic correlation between the mass of the $tZ$ system and the $t\bar{t}Z$ system  is a unique feature of signal events in this case.

\subsection{Final States with one Lepton and Missing Energy}

Final states involving missing energy can be more difficult to reconstruct compared to the leptonic $Z$ channel from the previous section, but come with the benefit of a large increase in the signal rate. Consider for instance the final state in which one of the two tops decays leptonically, while the $Z$ always undergoes a hadronic decay. The branching ratio to this final state is  $2 \times {\rm BR}(t\rightarrow \nu l b) \times {\rm BR}(t\rightarrow j j b) \times {\rm BR}(Z\rightarrow j j) = 2 \times 2/9 \times 2/3 \times 0.7 = 0.2$. Compared to the leptonic $Z$ decays with the branching ratio of 0.026, we see that requiring only one lepton and missing energy instead results in almost 8 times bigger signal event yield. The $1l + \MET$ final state, although it results in a higher signal yield, also suffers from much bigger backgrounds, the most dominant of which are SM $t\bar{t}$+jets and $W$+jets. 

Note that the search discussed in this section is significantly different from the standard searches for Super-summetry (SUSY) in the lepton + jets + $\MET$ channel. First, the search for $t\bar{t}Z$ resonances relies on the resonant structure of the final state which is different from the standard SUSY searches. Second, the kinematic regime we are trying to probe is different from the SUSY searches, making the existing results on lepton +jets +$\MET$ largely obsolete in our case.

We begin the event pre-selection by requiring presence of two $R=1.5$ jets with $p_T > 100 \GeV$, as the hadronic top and $Z$ candidates. In addition, we also require presence of the leptonic top candidate which we define as the system of a $b$-tagged jet with $p_T^b > 50 \GeV$, within $\Delta R_{bl } < 1.5$ from the hardest lepton in the event (with $p_T^l > 20 \GeV$), as well as $\MET > 50 \GeV$. If there are more than one $b$-tagged jets with $\Delta R_{bl } < 1.5$, we assign to the ``leptonic top'' the $b$-jet with the smallest $\Delta R_{bl }$.  Note that since we expect the signal events to be highly boosted, and we expect only one neutrino in the final state, the approximation of $\eta_\MET = \eta_l$ suffices to reconstruct the mass of the leptonically decaying top quark.

\begin{figure}[!]
\begin{center}
 \includegraphics[width=0.4\textwidth]{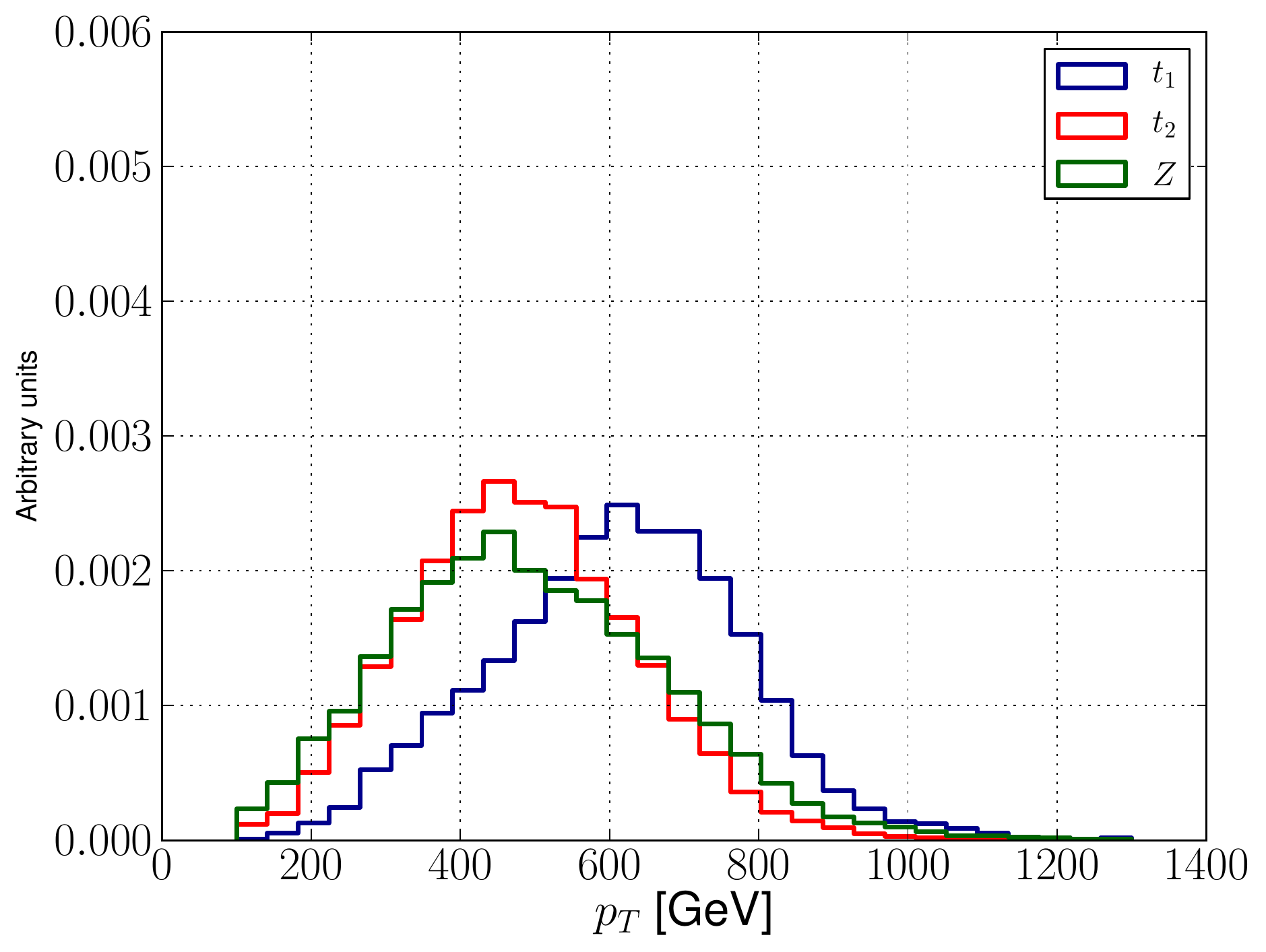}
  \includegraphics[width=0.4\textwidth]{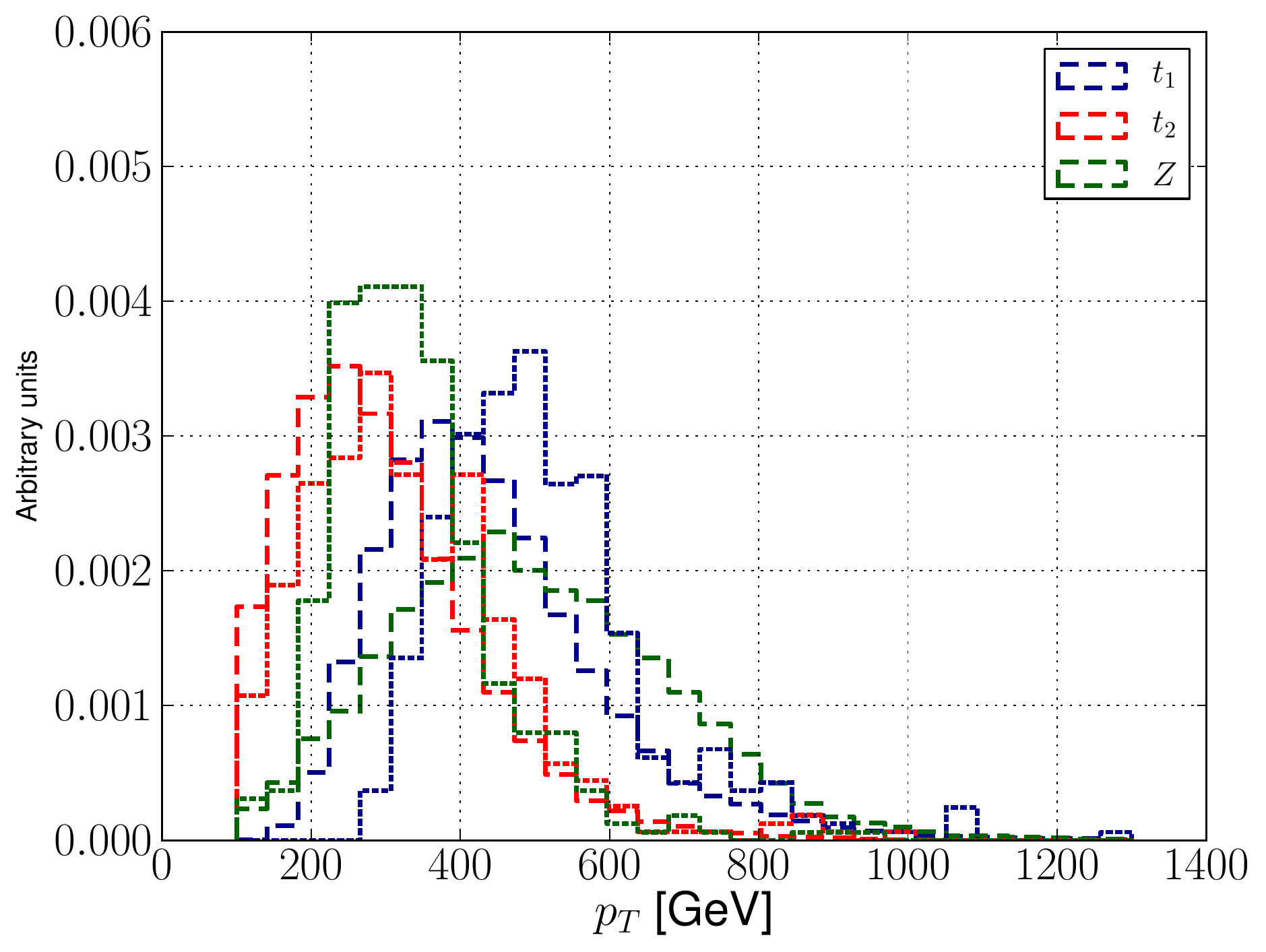}
    \includegraphics[width=0.4\textwidth]{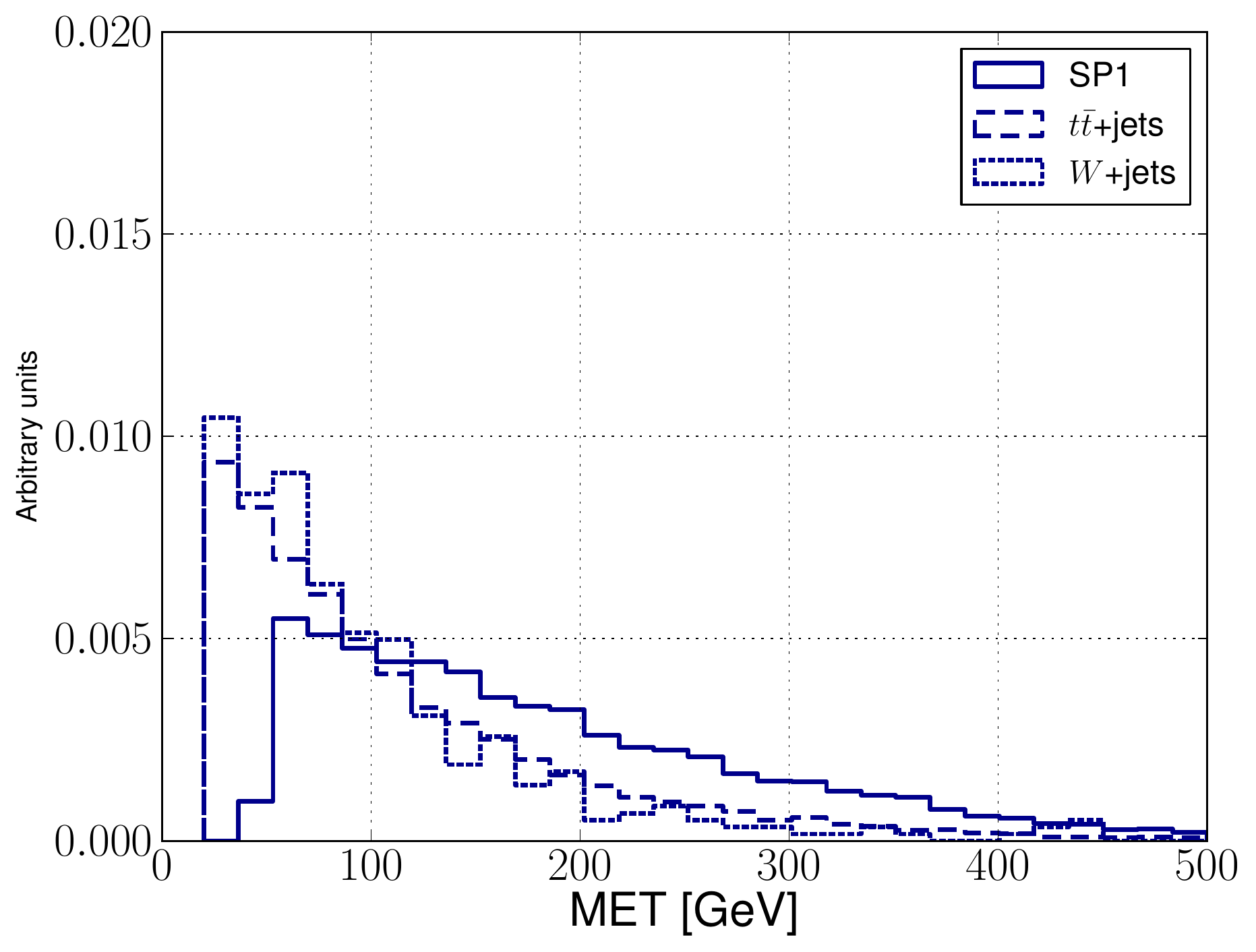}
        \includegraphics[width=0.4\textwidth]{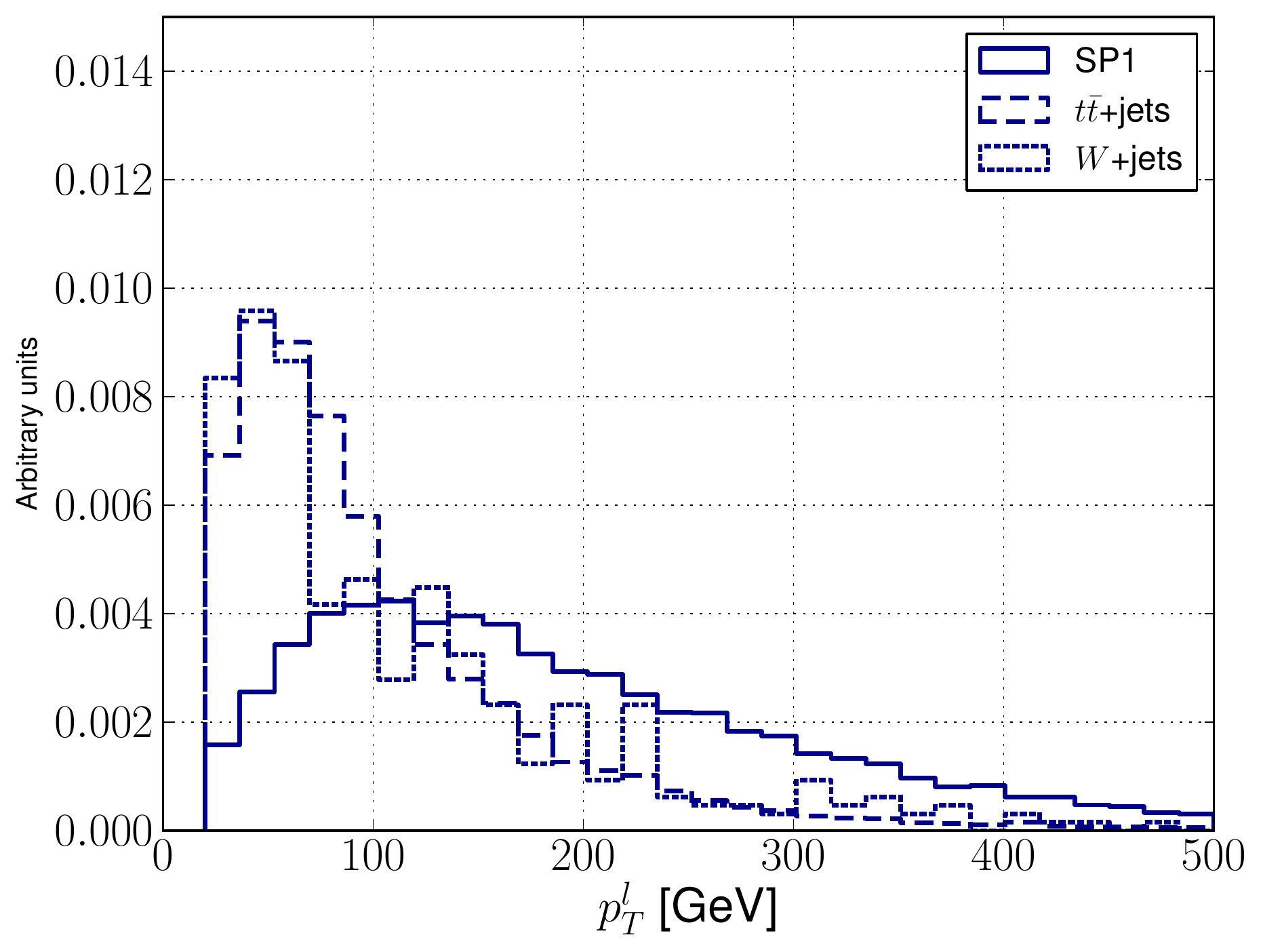}
 \caption{Distributions of kinematic observables for signal (solid lines) and background (dashed and dotted lines) events, for the $1l$+$\MET$ channel. For the purpose of illustration, we show the events generated for the SP1 model point while the background consists of $t\bar{t}$+jets and $W$+jets events. Labels $t_{1,2}$ refer to the hardest and second hardest $R=1.5$ jet in the event which is determined as the top candidate, and $Z$ represents the fat jet which we labeled as a $Z$ candidate. The events assume no pileup, detector simulation top or $Z$ tagging. All distributions are normalized to unit area.}
\label{fig:kinematics22}
\end{center}
\end{figure}

In order to determine which reconstructed object is the hardest top quark, the second hardest top quark and the  $Z$ boson, we employ a simple categorization scheme. If the leptonic top candidate $p_T$ is higher than the transverse momentum of the two hardest fat jets, we assign the leptonic top candidate to the hardest top quark in the event. The two remaining fat jets are categorized based on their mass asymmetry
$$
	\Delta m = \left| \frac{m_{\rm fj} - m_t}{m_{\rm fj} + m_t}\right|,
$$
where we assign to the second hardest top, the fat jet with a smaller $\Delta m$, while we assign to the $Z$ candidate, the remaining fat jet.
In the converse scenario where the leptonic top candidate is not the highest $p_T$ object in the event, we assign it to the second hardest top, and repeat the procedure of assigning the remaining fat jets to the hardest top quark and the $Z$ boson. 

\begin{figure}[!]
\begin{center}
 \includegraphics[width=0.4\textwidth]{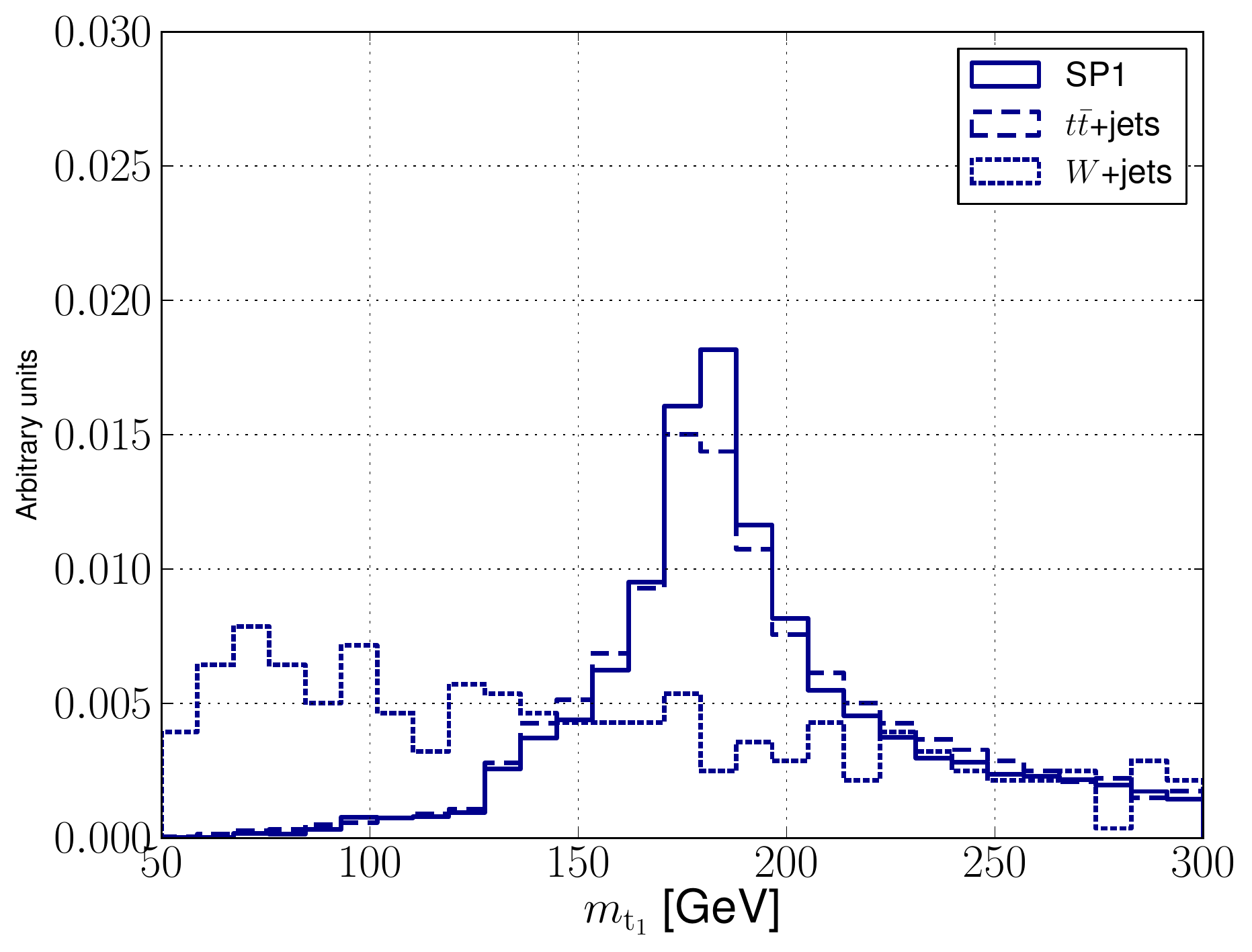}
  \includegraphics[width=0.4\textwidth]{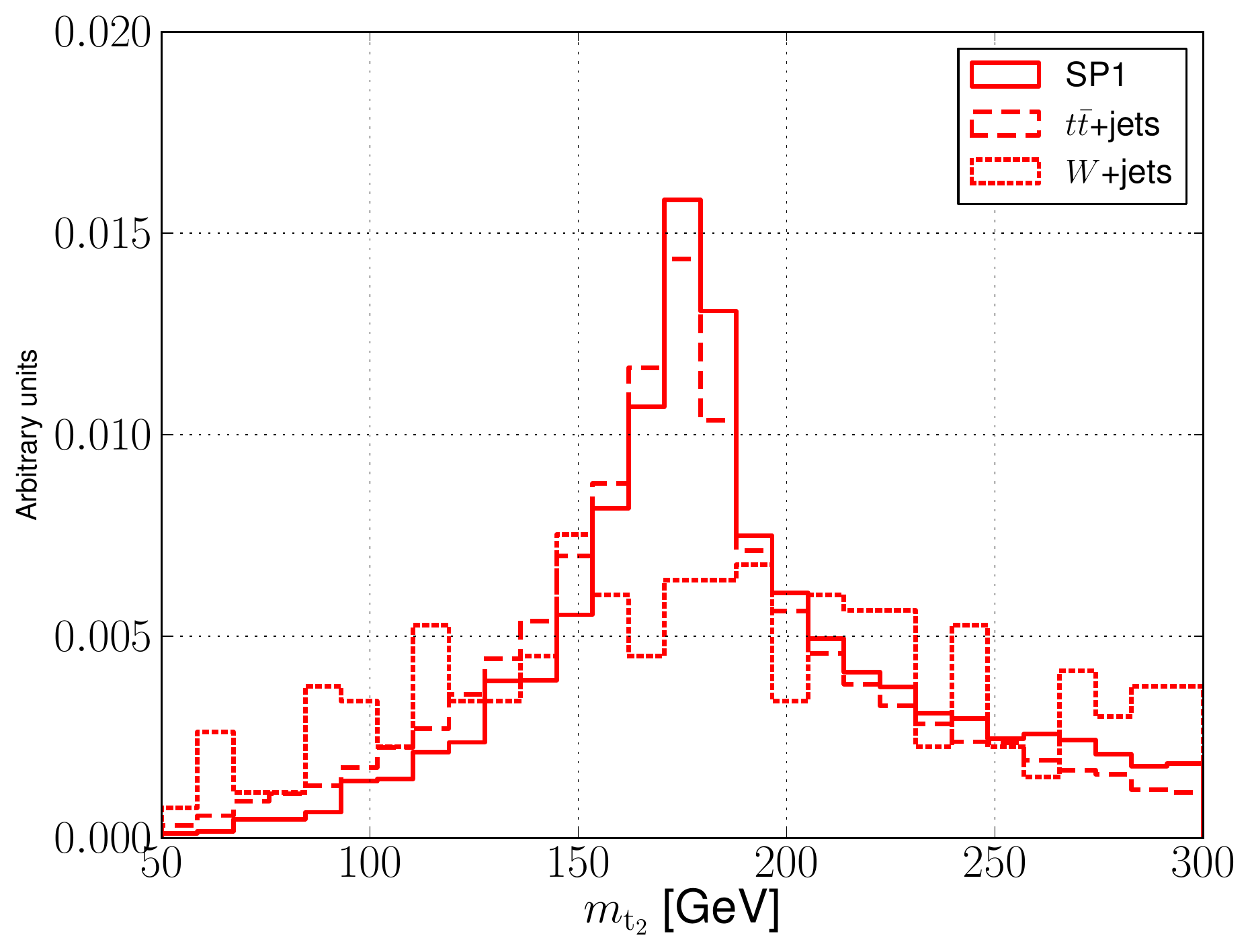}
    \includegraphics[width=0.4\textwidth]{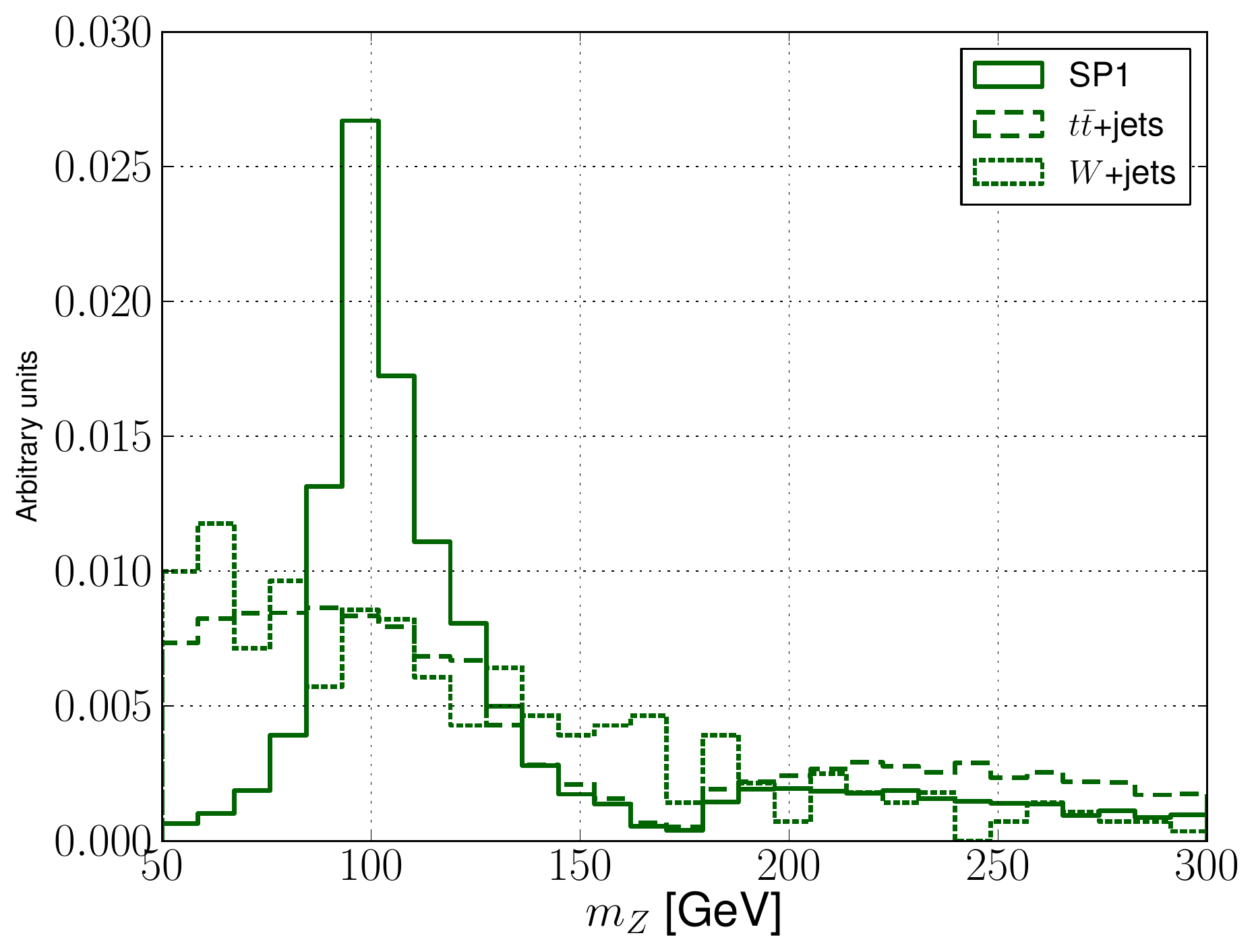}
        \includegraphics[width=0.4\textwidth]{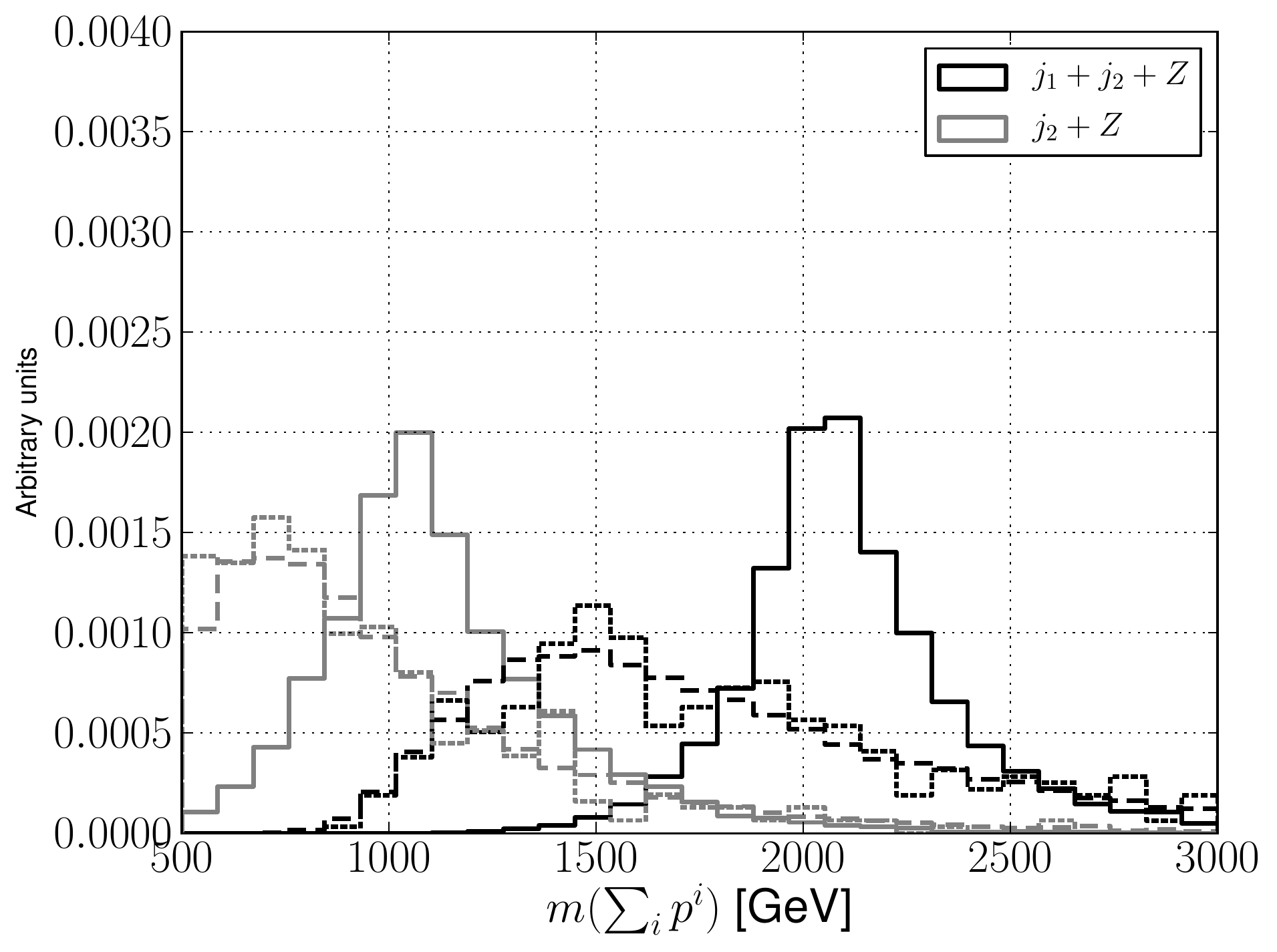}
 \caption{Distributions of kinematic observables for signal (solid lines) and background (dashed and dotted lines) events, for the $1l$+$\MET$ channel. For the purpose of illustration, we show the events generated for the SP1 model point while the background consists of $t\bar{t}$+jets and $W$+jets events. Labels $t_{1,2}$ refer to the hardest and second hardest $R=1.5$ jet in the event which is determined as the top candidate, and $Z$ represents the fat jet which we labeled as a $Z$ candidate. The events assume no pileup, detector simulation top or $Z$ tagging. All distributions are normalized to unit area.}
\label{fig:kinematics23}
\end{center}
\end{figure}

Figures~\ref{fig:kinematics22} and ~\ref{fig:kinematics23} show several examples of kinematic distributions relevant for the $1l$+$\MET$ channel. The first three panels show the invariant mass distribution of the hardest and second hardest top candidate object, as well as the $Z$ candidate. The plots illustrate well the performance of the procedure for determining whether the  fat jet is a candidate for a top quark or a $Z$ boson based on the $\Delta m$ criterion. The fourth panel illustrates the invariant mass distribution of the $t_2 Z$ system and the $t_1 t_2 Z$ system, where we show that the approximation of $\eta_\MET = \eta_l$ is sufficient to reconstruct the resonance masses. The remaining panels show the $p_T$ distributions of various reconstructed objects in the, where the signal events are in all cases characterized by a harder spectrum compared to the SM backgrounds. 

Since the signal events we are interested in are characterized by decays of $O(1 \TeV)$ objects, we require the hardest and the second hardest top candidates to have somewhat higher $p_T^{t_{1,2}} > 500, 400 \GeV$ respectively, while we require that the $Z$ candidate (which in the signal events should appear at the end of the decay chain) have $p_T > 300 \GeV$. Note that since the overall branching ratio is significantly higher compared to the $Z\rightarrow l^+ l^-$ channel,  we can utilize a higher $p_T$ cut on the reconstructed objects in order to suppress the background channels more efficiently, without sacrificing too much of the signal events.  Furthermore, in order to efficiently suppress the SM background with low missing energy, we also require $\MET > 100 \GeV$.  Finally, we tag the hadronic top and $Z$ jets using the CMS benchmark points described in Section~\ref{sec:pheno}. 

Table~\ref{tab:cutflow2} shows an example cutflow for the two benchmark model points we consider throughout the paper. The $W$+jets background is efficiently suppressed by, both the requirements on the boosted final state event topology and top tagging, where we find that each reduces the background contribution by an order of magnitude in cross section. We find that the LHC13 will be sensitive to the benchmark points we consider in the lepton +$\MET$ channel with a minimum of $\sim 300 \fb^{-1}$ of integrated luminosity, with the reach being inferior to the $Z\rightarrow l^+ l^-$ channel.

The somewhat inferior performance of the lepton + $\MET$ channel compared to $Z\rightarrow l^+ l^-$ is not unexpected. In case of the $Z\rightarrow l^+ l^-$, the double $b$-tag was crucial in suppressing the SM backgrounds, whereas in this case, the dominant background contribution already contains 2 real top quarks, and the $t\bar{t}$ rejection power comes mainly from boosted $Z$ tagging.  As boosted boson tagging is in general inferior to boosted top tagging, the background rejection power is hence limited compared to the $Z\rightarrow l^+ l^-$ case. 

\begin{table}[htb]
\begin{center}

\begin{tabular}{cccccc}

$(Z\rightarrow jj), 1l, \MET$& $\sigma$(SP1) & $\sigma$(SP2) & $\sigma$(SP3) & $\sigma$($t\bar{t}$+jets) & $\sigma$($W$+jets)  \\
\hline \hline
Preselection & 0.99 & 0.99 & 0.99 & 197 & 2.0 \\
$p_T^{t_{1,2}} > 500, 400  \GeV$ &0.57  & 0.56 &0.56  & 23 & 0.45 \\
$\MET > 100 \GeV$ & 0.46 & 0.46 & 0.46 & 18 &  0.23\\
$p_T^Z > 300 \GeV$ & 0.38 & 0.37 &  0.37 &10 & 0.14 \\
CMS top tag & 0.19 & 0.18 & 0.19 & 4.8 & $< 0.01 $\\
CMS $Z$ tag & 0.094 & 0.091 & 0.094 & 0.14 & $<$ 0.01 \\
\hline
$m_{23} > 800 \GeV$ & 0.088 & 0.087 & 0.087 & 0.13 &  $<$ 0.01 \\
$m_{123} > 1.8 \TeV$ &0.086 &0.084 & 0.086 & 0.12 &  $<$ 0.01 \\
\hline \hline
$S/B$& 0.72 & 0.72 & 0.72 &  \\
$S/\sqrt{B} (100  \fb^{-1})$& 2.5 & 2.5 & 2.5  & &  \\
$S/\sqrt{B} (300  \fb^{-1})$& 4.3 & 4.3 & 4.3 & &  \\
\hline
\end{tabular}
\end{center}
\caption{Example cutflow for channels with 1 hard lepton and missing energy. All samples assume a $H_T > 800 \GeV$ cut at the event generator level.  All cross section values are in fb. We use conservative $K$-factors of 2 and 1.3 respectively for the $t\bar{t}$ and $W$+jets background.}
\label{tab:cutflow2}
\end{table}

\subsection{Other Channels} \label{sec:otherchannels}

Final states other than two boosted jets and two leptons can be utilized in searches for $\rho_0 \rightarrow t ~T_{f_1}$ events. Among the final states containing a $Z$ boson there are:
\begin{itemize}
\item  $2l^+ 2l^- + 2b+{\rm MET}$: The final state is characterized with a high lepton multiplicity and hence relatively ``clean''. However, the channel is suppressed by ${\rm BR}(Z\rightarrow l^+ l^-)\times {\rm BR}(t\rightarrow l\nu b)^2 = 0.07 \times 0.11^2 = 8 \times 10^{-4}$, making it an unlikely discovery channel. This channel could, however, be significant in the follow-up studies at the high luminosity LHC.

\item $2\, {\rm fj} + \MET$: This final state could have good sensitivity for the discovery of an excess. As in case of $Z\rightarrow l^+ l^-$, the channel with large $\MET$ in addition to two top tagged boosted jets does not suffer from large backgrounds except for $Z$+jets, but benefits from a significantly higher signal event rate due to ${\rm BR}(Z\rightarrow \nu \bar{\nu}) \approx 22 \%$, and future studies should certainly consider this final state. Fully reconstructing the event and the resonance mass is significantly more difficult in this case due to the more complex composition of $\MET$. However, use of kinematic edge or transverse variables could provide useful information on the heavy particle masses.

\item {$ 3 \,{\rm fj}$}: The fully hadronic channel has the highest branching ratio ($\approx 30 \%$), but is accompanied by large backgrounds from SM multi-jet, $t\bar{t}$ and $W/Z$+jet processes, making it unlikely to be the discovery channel.

\end{itemize}

\section{Summary and Discussions}\label{sec:concl}

Past LHC searches for neutral vector resonances have mainly focused on two body resonance decays. Motivated by the absence of any new physics signal in resonance searches, as well as the fact that resonance mass limits are already well in the TeV range, we investigate the possibility that LHC is not discovering new resonances due to low resonance decay branching ratios into Standard Model two body final states. 
We argue that new neutral vector resonances can dominantly exhibit more complicated decay patterns,  where we focused mainly on the novel $t\bar{t}Z$ resonance decay channel. For the purpose of illustration, we explicitly showed how a dominant $\rho_0 \rightarrow t\bar{t}Z$ final state can appear in parameter regions of the two site composite Higgs model, where we use the 2-site Composite Higgs Model for concreteness. We demonstrate that in the context of CHM the decays $\rho \rightarrow t\, T_{f1}$ can dominate even when the $\rho_0 \rightarrow T_{f1} T_{f1}$ channel is kinematically open.

We identify specific model benchmark points which yield dominant BR($\rho_0 \rightarrow t\bar{t}Z$), while not being excluded by current experimental constraints. Note that our choice of model represents only one of many examples of a resonance search in three body final state states. 

The phenomenology of the $\rho_0 \rightarrow t\bar{t}Z$ decays is characterized by interesting final state kinematic configurations. If both the vector resonance and the top partner have masses of $O(1 \TeV)$, the final state consists of three heavy SM particle which are highly boosted and hence relatively simple to reconstruct. We study two particular scenarios. First, we consider final states where the $Z$ boson decays leptonically, while the $t$ and $\bar{t}$ decay hadronically. Second, we study the final states in which the $Z$ boson decays into jets, while either $t$ or $\bar{t}$ decay leptonically, but not both. 

We find that the $Z\rightarrow l^+ l^-$ scenario shows promising prospects for exploring the $t\bar{t}Z$ resonance decays. Our results indicate that the benchmark model points we consider could be discovered at LHC13 with as little as $30 \fb^{-1}$ of integrated luminosity. The powerful reach of the $Z\rightarrow l^+ l^-$ is mainly due to the low SM backgrounds, which can further efficiently be suppressed using boosted top tagging techniques, as well as by exploiting the characteristic doubly resonant structure in the invariant mass of two and three boosted objects. Searches for $t\bar{t} Z$ resonances in final states with one jets, one lepton and missing energy tend to suffer from more complicated and less reducible backgrounds and hence do not show sufficient sensitivity to discover the benchmark model points at the LHC in the near future. We also find that much about the sensitivity of before-mentioned searches depends on the  performance of top/$Z$ tagging algorithms.

Future studies of the $t\bar{t}Z$ final state could benefit from considerations of final states which contain two boosted hadronic tops, large missing energy and no  hard, isolated leptons. Even though characterized by more difficult resonance mass reconstruction, such final states suffer from relatively low SM backgrounds, as well as intrinsically higher signal rates,  and could hence prove to be useful channels for discovery of $t\bar{t}Z$. Furthermore, future studies would benefit from inclusion of detector effects, more realistic boosted object tagging algorithms,  as well as studies of high pileup environment on the results. 

Our results are generic enough that they can be applied to other three body resonance searches, such as $t\bar{t}h$ final states. The phenomenology of this final state, however would be somewhat different. Cases in which the search would, say, trigger on a single hard lepton in the event, would be similar to the $t\bar{t}Z$ scenario in which $Z$ boson decays hadronically, with the additional benefit that the Higgs tagging algorithm can exploit the presence of two $b$-jets inside the fat jet, and would hence surely perform better than the hadronic $Z$ search we discuss in this paper. The fully hadronic scenario could also yield useful results, as the overall signal rate is higher, albeit at the expense of large multi-jet backgrounds. Another generalization of our result could be in searches for charged vector resonances, where similar three body final states could be considered.


\section*{Acknowledgements}
We would like to thank Minho Son for useful discussions.
TF and BJ are grateful to the Mainz Institute for Theoretical Physics (MITP) for its hospitality and its partial support during the initial stages of this project.
This work was supported by the National Research Foundation of Korea (NRF) grant funded by the Korea government (MEST) (No. NRF-2015R1A2A1A15052408),
and by the Basic Science Research Program through NRF funded by the ministry of Education, Science and Technology (No. 2013R1A1A1062597)  and by IBS under the project code, IBS-R018-D1. 
SL was also supported in part by the Korean Research Foundation (KRF) through the Korea-CERN collaboration program (NRF-2016R1D1A3B01010529). MB is supported by the MOVE-IN Louvain Cofund grant.

\appendix
\section{Two -site Discrete Composite Higgs Model }\label{app:1}
In this Appendix, we give the details on the two-site composite Higgs model \cite{Panico:2011pw} for which we define the three benchmark points detailed in Appendix \ref{app:2}, which are used for our study in section~\ref{sec:pheno}. The Lagrangian of the model have been discussed in some detail before in the Appendix of Ref.~\cite{Low:2015uha} and (for a very similar effective parametrization) Ref. \cite{Greco:2014aza}.  We summarize the results again in this Appendix for the convenience of the reader and to have a complete definition of the model used for our simulations.

The leading order Lagrangian of the two-site composite Higgs model is 
\begin{equation}\label{eq:lstruc}
\mathcal{L}_{0}= \mathcal{L}_{nl\sigma}+\mathcal{L}_{gauge}+\mathcal{L}_{fermion}
\end{equation}
where the non-linear $\sigma$-model term is given by 
\begin{equation}\label{eq:nlsigma}
\mathcal{L}_{nl\sigma} =
\frac{f^2}{4}\mathrm{Tr}[(D_\mu \mathcal{U})^T D^\mu \mathcal{U}] \qquad \text{with} ~\mathcal{U} = \exp\left( i \frac{\sqrt{2}h^i\hat{T^i}_{IJ}}{f} \right),
\quad i=1,2,3,4,
\end{equation}
where $\hat{T^i}_{IJ}$ are the  4 broken generators of $SO(5)_V$ in the fundamental representation, which can be written as 
\begin{equation}
\hat{T^i}_{IJ}= -\frac{i}{\sqrt{2}}\left(\delta_I^i\delta^5_J-\delta_J^i\delta^5_I\right)\quad \quad \text{where} ~I,J= \{1,\cdots,5\}.
\end{equation}
These broken generators are associated with the Goldstones, $h^i$, which live in $\mathbf{4}$ of $SO(4)$, and  can be related to the two complex Higgs doublet components, $h_u$ and $h_d$ as, 
\begin{equation}
	H=\left(
	\begin{array}{c}
	h_u \\
	h_d \\
	\end{array}
	\right)=\frac{1}{\sqrt{2}}\left(
	\begin{array}{c}
	i h_1+h_2 \\
	h_4-i h_3 \\
	\end{array}
	\right).
\end{equation}
In unitary gauge the Goldstone matrix takes a very simple form and becomes
\begin{equation}
	\mathcal{U}=\left(
	\begin{array}{ccc}
	\mathds{1}_{3} & \vec{0}&\vec{0} \\
	\vec{0} & \cos \frac{v}{f} & \sin \frac{v}{f} \\
	\vec{0} & -\sin \frac{v}{f} & \cos \frac{v}{f} \\
	\end{array}
	\right)
\end{equation}
where $	\mathds{1}_{3}$ is the $3\times3$ identity matrix and $\vec{0}$ is the 3-dimensional null vector.
 The covariant derivative, which gives all the interactions of SM gauge bosons and other heavy vector resonances with the Goldstones, transforms $\mathcal{U}$ as
\begin{equation}
D_\mu \mathcal{U}= \partial_\mu \mathcal{U} - i \hat{g} \hat{A}_\mu^a T_L^a \mathcal{U} -i \hat{g}' \hat{B}_\mu T_R^3 \mathcal{U} + i \hat{g}_\rho \mathcal{U} \hat{\rho}_\mu^b T^b.
\end{equation}

 The elementary gauge fields are labelled by $\hat{A}_\mu^a$ and $\hat{B}_{\mu}$ while the composite counterparts with $\hat{\rho}_\mu^b$ . The index $a$ runs over $1,2,3$ and $b$ spans $\{1,\cdots,6\}$. $T^a$'s are the 6 unbroken generators which form an  $SO(4)$ algebra and are written in terms of the $SU(2)_L\times SU(2)_R$ notation in terms of  $4\times4$ generators, $t^{\alpha }_{L,R}$ . 
\begin{equation}
T^{a}= \begin{Bmatrix}
T^{\alpha}_L=\left[
\begin{array}{cc}
t^{\alpha }_L & 0 \\
0 & 0 \\
\end{array}
\right],
\quad 
T^{\alpha}_R=\left[
\begin{array}{cc}
t^{\alpha }_R & 0 \\
0 & 0 \\
\end{array}
\right] 
\end{Bmatrix}  
\qquad , \alpha={1,2,3}.
\end{equation}
The  first 3 generators of $T^b$ are identified with $T^{\alpha}_L$ and the remaining with $T^{\alpha}_R$. 

\subsection{Gauge sector}

The vector part of the Lagrangian can be written as 
\begin{equation}
\mathcal{L}_{gauge}= \mathcal{L}^{elementary}+\mathcal{L}^{composite}
\end{equation}
where
\begin{eqnarray}
&\mathcal{L}^{elementary}=& -\frac{1}{4} \hat{W}^a_{\mu\nu}{}^2 -\frac{1}{4} \hat{B}_{\mu\nu}{}^2 \nonumber \\  &\mathcal{L}^{composite} =&-\frac{1}{4} \hat{\rho}^b_{\mu\nu}{}^2
\label{eq:laggauge}
\end{eqnarray}

The $\hat{W}$, $\hat{B}$, are the elementary (``first site'') $SU(2)_L$ and U(1)$_Y$ gauge bosons with their associated gauge couplings $\hat{g}$ and  $\hat{g}'$ . The ``second site" contains the composite $\hat{\rho}$ gauge fields with  $g_\rho$ as the strong gauge coupling. The $\hat{\rho}^b$ field can be rewritten as an SU(2)$_L$ field $\hat{\rho}_L$ and an SU(2)$_R$ field $\hat{\rho}_R$.  It should be noted that the ``hat" indicates that all these fields and couplings are defined before the symmetry breaking.


The above equations contain mass-term mixings for the elementary $\hat{W}$,~$\hat{B}$ and the composite $\hat{\rho}$ ~fields. We can rotate it to the physical basis by diagonalizing the mass matrices which arise from the Eq.~(\ref{eq:nlsigma}).  The mass terms can be written as 
\begin{equation}
\mathcal{L}_{mass}= V^{+} M^2_{\pm} V^{-} + \frac{1}{2} V^0 M_0^2 V^0
\end{equation}
where $V^{\pm}= (V^1\pm i V^2)/\sqrt{2},~~~ V^{1,2}=\{\hat{W}^{1,2},\hat{\rho}^{1,2}_L,\hat{\rho}^{1,2}_R\}$ and $ V^0=\{\hat{W}^3,\hat{\rho}_L^3,\hat{B}^3,\hat{\rho}_R^3\}$. The charged and neutral mass matrices are given by 
\begin{equation}
M^2_{\pm}=\left(
\begin{array}{ccc}
\frac{1}{2} f^2 \hat{g}^2 & -\frac{1}{4} f^2 \hat{g} g_{\rho} \left(c_{\theta
}+1\right) & \frac{1}{4} f^2 \hat{g} g_{\rho} \left(c_{\theta }-1\right) \\
-\frac{1}{4} f^2 \hat{g} g_{\rho} \left(c_{\theta }+1\right) & \frac{1}{2} f^2
g_{\rho}{}^2 & 0 \\
\frac{1}{4} f^2 \hat{g} g_{\rho} \left(c_{\theta }-1\right) & 0 & \frac{1}{2} f^2
g_{\rho}{}^2 \\
\end{array}
\right)
\end{equation} 
\begin{equation}
 M_0^2=\left(
 \begin{array}{cccc}
 \frac{1}{2} f^2 \hat{g_y}^2 & 0 & -\frac{1}{4} f^2 g_{\rho} \hat{g_y} \left(c_{\theta
 }+1\right) & \frac{1}{4} f^2 g_{\rho} \hat{g_y} \left(c_{\theta }-1\right) \\
 0 & \frac{1}{2} f^2 \hat{g}^2 & \frac{1}{4} f^2 \hat{g} g_{\rho} \left(c_{\theta
 }-1\right) & -\frac{1}{4} f^2 \hat{g} g_{\rho} \left(c_{\theta }+1\right) \\
 -\frac{1}{4} f^2 g_{\rho} \hat{g_y} \left(c_{\theta }+1\right) & \frac{1}{4} f^2
 \hat{g} g_{\rho} \left(c_{\theta }-1\right) & \frac{1}{2} f^2 g_{\rho}{}^2
 & 0 \\
 \frac{1}{4} f^2 g_{\rho} \hat{g_y} \left(c_{\theta }-1\right) & -\frac{1}{4} f^2
 \hat{g} g_{\rho} \left(c_{\theta }+1\right) & 0 & \frac{1}{2} f^2 \hat{g_{\rho
 	}}{}^2 \\
 	\end{array}
 	\right)
\end{equation}

where, $c_{\theta}= \cos v/f$.

The mass matrices can be diagonalized analytically  but the exact expressions of its eigenvectors and eigenstates are lengthy. Thus in the following we only give the explicit expressions for the eigenvalues to leading order in $v/f$.  We obtain the exact results by numerically diagonalizing the mass matrices.
The physical masses in the small $v/f$ limit are: 
\begin{equation}\label{eq:spectrum}
\begin{aligned}
&\quad m_W^2 =\frac{ v^2 \hat{g}^2 g_{\rho}^2}{4 \left(g_{\rho}^2+\hat{g}^2\right)},\\
&\quad m_Z^2 =\frac{1}{4} v^2 g_{\rho}^2 \left(\frac{\hat{g_y}^2}{\hat{g_y}^2+g_{\rho}^2}+\frac{\hat{g}^2}{g_{\rho}^2+\hat{g}^2}\right), \\
\mathbf{3}_0 &:m_{\rho_{0,\pm}}^2 =\frac{1}{2} f^2 \left(g_{\rho}^2+\hat{g}^2\right)-\frac{\hat{g}^2 v^2 g_{\rho}^2}{4 \left(g_{\rho}^2+\hat{g}^2\right)},\\
\mathbf{1}_0 &:m_{\rho_B}^2 =\frac{1}{2} f^2 \left(\hat{g_y}^2+g_{\rho}^2\right)-\frac{v^2 g_{\rho}^2\hat{g_y}^2}{4 \left(\hat{g_y}^2+g_{\rho}^2\right)}, \\
\mathbf{1}_{\pm} &:m_{\rho_C}^2 =\frac{1}{2} f^2 g_{\rho}^2.		
\end{aligned}
\end{equation}
At this order, SM electroweak couplings are related to 2-site model gauge couplings as, 
\begin{subequations}
	\begin{align}
	\frac{1}{g^2}  &= \frac{1}{\hat{g}^2} + \frac{1}{g_{\rho}^2}, \\
	\frac{1}{g_y^2} &= \frac{1}{\hat{g}'^2} + \frac{1}{g_{\rho}^2}.
	\end{align}
\end{subequations}

The above relations show that the values $\hat{g},\hat{g}'$ have to be chosen accordingly for a given value of $g_{\rho}$ in order to reproduce the SM masses of the $W$ and $Z$ boson and the SM couplings $g,g_y$. To perform the numerical diagonalization (beyond leading order in $v/f$) we fix $\hat{g}, \hat{g}'$ by demanding the lowest charged eigenvalue to be $M_{W,physical}$ and second lightest e-value in the neutral sector by the Z boson mass, $M_{Z, physical}$.  


\subsection{Matter content}

We implement the partially composite scenario, where we work along the lines of ~\cite{Backovic:2014uma}. The top quark doublet $(t_L,b_L)$ and also the singlet, $t_R$ get a sizable degree of compositesness. For this purpose, a Dirac 5-plet top partner multiplet is introduced which transforms non-linearly under $SO(5)$.  To ensure that we get the correct hypercharge for the fermions, an extra $U(1)_X$ global symmetry is introduced. Note that as the Goldstones are not charged under $U(1)_X$, there are no implications on the gauge sector. 
The top partner 5-plet 
\begin{equation}
\Psi= \left(
\begin{array}{c}
\Psi _4 \\
\Psi _1\\
\end{array}
\right)=\left(
\begin{array}{c}
i B'-i X_{\frac{5}{3}} \\
X_{\frac{5}{3}}+B' \\
i T+i X_{\frac{2}{3}} \\
X_{\frac{2}{3}}-T \\
\tilde{T} \\
\end{array}
\right) 
\end{equation}
with $U(1)_X$ charge 2/3, decomposes into a 4-plet (bi-doublet) $((X-{5/3},X_{/23}),(T,B'))$ and a singlet $\tilde{T}$ under $SO(4)$ ($\simeq SU(2)_L\times SU(2)_R$), where $X_{5/3}$ has electric charge $5/3$, the states $X_{2/3} , T, \tilde{T}$ have charge $2/3$, and $B'$ has charge $-1/3$. 
  
The elementary third generation quarks are embedded as an incomplete 5-plet - which transforms linearly under $SO(5)$. 
\begin{equation}
q_L=\frac{1}{\sqrt{2}}\left(
\begin{array}{c}
i b'_L \\
b'_L \\
i t'_L \\
-t'_L \\
0 \\
\end{array}
\right) \quad\quad t_R=\left(
\begin{array}{c}
0 \\
0 \\
0 \\
0 \\
t'_R \\
\end{array}
\right)
\end{equation}

The top-sector Lagrangian, written in the gauge eigenbasis is 
\begin{equation}\label{eq:ltop}
\mathcal{L}_{\rm top} = i \overline{q}_L  \slashed{D} q_L +i \overline{t}_R  \slashed{D} t_R +i \overline{\Psi}_4 (\slashed{D}- M_4)  \Psi_4+i \overline{\Psi}_1 (\slashed{D}- M_1)  \Psi_1+ y_L f \overline{q}_L(\mathcal{U} \Psi)+y_R f \overline{t}_R(\mathcal{U} \Psi)+ h.c \cdots
\end{equation}

The gauge eigenstates mix owing to EWSB and the resulting mass eigenstates are: (i) two states:  SM bottom, $b$ and it's partner, $B$ with EM charge $-1/3$, (ii) four states:  comprising the SM top, $t$ accompanied by the top partners, $T_{f1},T_{f2}, T_s$ with charge $2/3$ and (iii) $X_{5/3}$ with an exotic charge $5/3$. The top sector mass terms can be written as 
\begin{equation}
 \mathcal{L}_{\rm top,m}= - \overline{\Psi} _{t,L}M^t \Psi _{t,R}+ -\overline{\Psi}
 	_{b,L}M^b \Psi _{b,R}-M_4 \overline{X}_{\frac{5}{3},L}X_{\frac{5}{3},R} + h.c \cdot
\end{equation}

where 

\begin{equation}
	M^t=\left(
	\begin{array}{cccc}
	0 & f \cos ^2\left(\frac{v}{2 f}\right) y_L & f \sin ^2\left(\frac{v}{2 f}\right) y_L &
	-\frac{f \sin \left(\frac{v}{f}\right) y_L}{\sqrt{2}} \\
	\frac{f \sin \left(\frac{v}{f}\right) y_R}{\sqrt{2}} & M_4 & 0 & 0 \\
	-\frac{f \sin \left(\frac{v}{f}\right) y_R}{\sqrt{2}} & 0 & M_4 & 0 \\
	f \cos \left(\frac{v}{f}\right) y_R & 0 & 0 & M_1 \\
	\end{array}
	\right) \quad  \text{and}   \quad M^b=\left(
	\begin{array}{cc}
	0 & y_L f \\
	0 & M_4 \\
	\end{array}
	\right).
\end{equation}

In the mass basis (i.e after mass diagonalization) the masses of the top and its partners in the lowest order of $(v/f)^2$ are given by 
\begin{equation}\label{eq:topspectrum}
\begin{aligned}
m_t &\simeq \frac{v}{\sqrt{2}}\frac{|M_1-M_4|}{f}\frac{y_L y_R f^2}{\sqrt{M_4^2+y_L^2f^2}\sqrt{M_1^2+y_R^2f^2}},\\
m_{T_{f,1} } &\simeq M_4,\\
m_{T_{f,2} } &\simeq \sqrt{M_4^2 + f^2y_L^2}, \\
m_{T_s} &\simeq \sqrt{M_1^2 + f^2y_R^2}, ,\\
m_{X_{5/3}} &= M_4, \\
m_B &= \sqrt{M_4^2 + f^2y_L^2}, 	
\end{aligned}
\end{equation}
where $T_{f,1}, T_{f,2}, T_s, X_{5/3},B$ label the top partner mass eigenstates.  For (long) explicit expressions of the transformation matrices from the gauge to the mass eigenbasis at leading order in $(v/f)^2$, we refer to  the Appendix of Ref.~\cite{Backovic:2014uma}. Again (like for the gauge sector mass diagoanlization), these leading order expressions are only given for orientation, and for our sample points in the next appendix as well as for our simulations, we diagonalize the mass matrices numerically.

\bigskip

The above described top-partner sector is of special relevance in extracting the couplings of vector resonances to 3rd family quarks, and thus to determine vector resonance branching ratios. The top sector is particularly cumbersome because in order to produce a sizable top mass (which is of the order of $v/\sqrt{2}$), both, left- and right handed pre-Yukawa $y_{L,R}$ couplings have to be present, which yields non-negligible mixing in both, the left- and the right-handed top-top-partner sector.

Analogs of the above embedding of tops and top partners can in principle be used to give mass to other quarks via partial compositeness (at the price of including numerous vector-like partners), but these are much simpler to parameterize: To obtain a quark mass $m_q\ll v/\sqrt{2}$, at most one pre-Yukawa coupling $\lambda^q_{L,R}$ can be sizable an lead to a non-negligible mixing. In this article we assume a dominant universal  mixing angle for the first and second family in the left-handed sector. This mixing can then be parameterized by a single mixing angle $\theta^q_L$ which appears in couplings through $s_{L,q}\equiv \sin(\theta^q_L)$.

\subsection{Vector resonance interactions}

From the Lagrangian (\ref{eq:nlsigma}), (\ref{eq:laggauge}), and  (\ref{eq:ltop}) we can determine the interactions of the heavy vector resonances in the mass eigenbasis. Due to the mixing of composite vector resonances with elementary vector resonances, the heavy vector resonances obtain direct couplings to standard model fermions (as well as gauge bosons). Quarks obtain a second contribution due to the mixing of elementary quarks with composite fermionic resonances. In more detail,  the cubic interactions involving the neutral heavy resonance, SM fermions and top partners are: 
\begin{eqnarray}
\mathcal{L}_{\rho} &=& i g_{\rho_{0} WW} \frac{m_W^2 }{g_{\rho} f^2} \left[( \partial_{\mu} W_{\nu}^{+} - \partial_{\nu} W_{\mu}^{+} ) W^{\mu -} \rho^{0 \nu}  + ( \partial_{\mu} \rho_{\nu}^{0}- \partial_{\nu} \rho_{\mu}^{0} )W^{\mu +} W^{\nu -} + h.c. \right]  \nonumber \\ 
&+&  m_Z ~g_{\rho}~g_{\rho_{0} Zh}  ~h \rho_{\mu}^{0} Z^{\mu} + \sum_{X,Y} \lambda_{h X_L Y_R} ~h X_L Y_R 
+ \sum_{X,Y} \left[   g_{\rho_{0} X Y}^{L}   X_{L} \slashed{\rho}_{0} \bar{Y_L} + g_{\rho_{0} X Y}^{R}   X_{R} \slashed{\rho}_{0} \bar{Y_R}    \right] \nonumber \\ 
&+&  \sum_{X,Y} \left[   g_{Z X Y}^{L}   X_{L} \slashed{Z} \bar{Y_L} + g_{Z X Y}^{R}   X_{R} \slashed{Z} \bar{Y_R}    \right] 
+ \sum_{X,Y} \left[   g_{W X Y}^{L}   X_{L} \slashed{W} \bar{Y_L} + g_{\rho_{0} X Y}^{R}   X_{R} \slashed{W} \bar{Y_R}    \right] \nonumber \\ 
&+& \sum_{q=1}^2   g_{\rho_{0} qq}^{L} ~  q_{L} \slashed{\rho}_{0} \bar{q_L}       
+  \sum_{l=1}^3   g_{\rho_{o} ll}^{L} ~  l_{L} \slashed{\rho}_{0} \bar{l_L}    \label{Eq:Leff}
\end{eqnarray}
where X,Y = ${t,b, T_{f1}, T_{f2}. T_{s}, X_{5/3}, B}$ , i.e  the top, bottom and its partners,  $q_L$ are the light quarks and $l_L$ are the leptons. 

Through these couplings, the cubic interactions listed in Lagrangian~(\ref{Eq:Leff}), dictate the production and decay of the neutral heavy resonances. 

Heavy vectors can be produced from (i) Drell-Yan: depends on the $\rho_{\pm} q \bar{q}$ coupling, and (ii) Vector boson fusion: however, VBF is very subdominant~\cite{Low:2015uha}. The produced neutral heavy vector decays into SM final states and also new exotic final states involving the top partner. It is possible to allow for single top partner associated with top quark to be the dominant final state. This top partner can decay into $bW$, $th$ and $tZ$ final states. 

The couplings of interest are $\rho_{\pm} q \bar{q}$, see~Eq.~(\ref{eq:rhoqq}), couplings to the light quarks($j$),  $g_{\rho^0 ff}$   and leptons($l$), top and bottom quarks, $g_{\rho^0 tt}$, $g_{\rho^0 bb}$ respectively. Interactions of the vector resonances with the top partners are encoded in the couplings,  $g_{\rho^0 tX}$ or $g_{\rho^0 Y X}$ where ${X,Y}={T_{f1},T_{f2}, T_s}$  and also $g_{\rho^0 X_{5/3} X{5/3}}$ for the $5/3$ charged top partner. Lastly, we have couplings from the bottom and its partner $g_{\rho^0 b B}$ and $g_{\rho^0 B B}$.  
The couplings relevant for the production and decay of the neutral vector resonance $\rho_0$ are\footnote{Again, we give analytic expressions to leading order in $v/f$ but use the full result from numerical diagonalization for our simulation.} 

\begin{equation}
\begin{aligned}
                 g^L_{\rho_0 q\bar{q}} &= -\frac{\hat{g}^2 }{g_{\rho }}\left(1-\frac{g_{\rho }^2 ~s_{L,q}^2}{\hat{g}^2}\right)\, , &\quad
                 g^R_{\rho_0 q\bar{q}} &=0\, ,\\
                 g^L_{\rho_0 T_{f,1}t} &=\frac{\epsilon^2  g_{\rho}}{3}\frac{s_y s_{L,t} s_{R,t}^2 \left(M_{T_s}^2+M_1(M_1-M_4)\right)}{(M_{T_s}^2-M_4^2)}\, ,&\quad
                 g^R_{\rho_0 T_{f,1}t} &= c_y~ s_{R,t}~ \frac{v}{f}~  \frac{g_{\rho}}{2 \sqrt{2}}  \frac{ M_1}{M_4}\, ,  \\
		g^L_{\rho_0 tt}&= \frac{\hat{g_y} s_y \left(3 f^2 y_L^2 + M_{B}^2\right)}{6 M_{B}}\, , &\quad 
		g^R_{\rho_0 tt}&=  \frac{2 \hat{g_y} s_y}{3}\, , \\		
		g^L_{\rho_0 bb}&= \frac{3g_{\rho} c_y s_{L,t}^2 \hat{g}_y s_y (1+ 3  s_{L,t}^2) }{6}\, ,  &\quad 
		g^L_{\rho_0 bB}&= \frac{ c_{L,t}s_{L,t} \hat{g}_y }{ s_y}\, ,  \\	
		g^L_{\rho_{0} t T_{f,2}}&= \hat{g}_y s_y s_{L,t} \frac{M_4}{M_{B}}\, ,  &\quad 
		g^R_{\rho_{0} t T_{f,2}}&= \frac{g_{\rho} c_y s_{R,t}}{\sqrt{2}}\left(1+ \frac{M_4(M_1-M_4)}{M_{B}^2}\right)\, ,  \\
		g^L_{\rho_{0} t T_{s}}&= \frac{\hat{g}_y s_y s_{L,t}  \frac{(M_1-M_4)}{\sqrt{2} M_{T,s}^2}} {1 - \frac{M_B^2+M_4 M_1}{M_{T_s}^2+M_4 M_1}}\, ,  &\quad 
		g^R_{\rho_{0} t T_{s}}&= 0\, ,  \\		
\end{aligned} 
\end{equation}

\begin{equation}
\begin{aligned}
g^{L,R}_{\rho_0 T_{f,1} T_{f,1}}&= g^{L,R}_{\rho_0 X_{5/3} X_{5/3}}= \frac{3 g_{\rho} c_y -4 \hat{g}_y s_y }{6}\, ,  &\quad
g^{L,R}_{\rho_0 T_{s} T_{s}}= \frac{2 g_y s_y}{3}\, , \\
g^{L}_{\rho_0 T_{f,2} T_{f,2}}&= \frac{ g_y s_y (M_B^2+3 M_4^2)}{6 M_B^2}\, , &\quad 
g^{R}_{\rho_0 T_{f,2} T_{f,2}}= \frac{3 g_{\rho} c_y +4 \hat{g}_y s_y }{6}\, , 
\end{aligned} 
\end{equation}
where the mixing angles in the gauge and top sector are 

\begin{equation}\label{eq:gmix}
\begin{aligned}
s_y= \frac{\hat{g}_y}{\sqrt{\hat{g}_y^2 + g_{\rho}^2}} \quad \text{and}  \quad c_y= \frac{g_{\rho}}{\sqrt{\hat{g}_y^2 +g_{\rho}^2}}\, , \\
s_2= \frac{\hat{g}}{\sqrt{\hat{g}^2 + g_{\rho}^2}} \quad \text{and}  \quad c_2= \frac{g_{\rho}}{\sqrt{\hat{g}^2 + g_{\rho}^2}}\, ,  
\end{aligned} 
\end{equation}

\begin{equation}\label{eq:tmix}
\begin{aligned}
s_{L,t}= \frac{y_L f}{\sqrt{y_L^2 f^2 +M_4^2}} \quad \text{and}  \quad c_{L,t}=\frac{M_4}{\sqrt{y_L^2 f^2 +M_4^2}}\, , \\
s_{R,t}= \frac{y_R f}{\sqrt{y_R^2 f^2 +M_1^2}} \quad \text{and} \quad c_{R,t}=\frac{M_1}{\sqrt{y_R^2 f^2 +M_1^2}}\, .
\end{aligned} 
\end{equation}

Here, ($M_1, M_4$) are  the single and fourplet masses  which enter composite sector for the top sector and $y_{L,R}$  are the left and right handed preyukawa terms ,responsible for mixing the elementary and composite fermionic sectors. 


The couplings of  responsible for the decay of $T_{f,1}$ to $b W_{\pm}$  and $h Z$ at the lowest order are given by

\begin{equation}
	\begin{aligned}
	         &g^R_{T_{f,1}tZ} =  s_y~ s_{R,t} ~\frac{v}{f}~~ \frac{g_{\rho}}{2 \sqrt{2}}  \frac{ M_1}{M_4}  \\
		&g^L_{T_{f1}bW_{\pm}}= -\eps^2 \frac{ s_{L,t} ~s_2 ~g_{\rho}}{\sqrt{2}}   \\
		&g^R_{T_{f1}tW_{\pm}}= \mathcal{O}(\eps^3), \\
		&\lambda^L_{h t_R T_{f1,L}} = - \frac{M_1 }{\sqrt{2} f} s_{R,t},   \\
		&\lambda^R_{h t_L T_{f1,R}} = \eps y_L \dfrac{(M_1 - M_4) }{M_4 M_B} \dfrac{(f^2 y_R^2 M_1 + M_4^2 (M_1+M_4))}{(M_4^2-M_{T_s}^2)}\, ,
	\end{aligned}
\end{equation}
where $\epsilon = v/f $, $s_w = \hat{g}_y/\sqrt{\hat{g}^2+\hat{g}_y^2}$,  $c_w = \hat{g}/\sqrt{\hat{g}^2+\hat{g}_y^2}$.

The couplings of the $T_{f1}$ are relevant for this article as in our phenomenological study we focus on production of the $\rho_0$ and subsequent decay into $T_{f1}\bar{t}$, and our benchmark points are chosen such that this decay is kinematically allowed, while other charge 2/3 top partners are heavier than the $\rho_0$. Expressions for the couplings of the other top partners to standard model gauge bosons and third generation quarks can be found in the Appendix of Ref.~\cite{Backovic:2014uma}. 
\section{Details of the Benchmark Models}\label{app:2}
In this article we focus on the production and the decay of the neutral heavy resonance $\rho_0$, and in particular the decay into a top and a top partner. For purpose of simulation, we choose three benchmark points in which a top partner ($T_{f1}$) is either slightly lighter, heavier, or of similar mass as $m_\rho/2$, such that $\rho_0$ can always decay into a top and its partner,  while the decay into a pair of top partners is either kinematically forbidden, suppressed, or allowed. As described in subsection.~\ref{sec:benchmark}, we fix the underlying model parameters to the values given in Eq.~(\ref{eq:BMfix}):
 \begin{equation}
 g_{\rho}= 3.5 ,~~ f=808 ~\text{GeV},~~ m_{\rho}= 2035 ~\text{GeV} ,~~ M1= 20 ~\text{TeV}, ~~s_{L,q}= 0.1\, , \nonumber
 \end{equation}
common to all three benchmark points, and in addition choose three sets of values for $(M_4,y_R)$ given in Table~\ref{tab:BMdiff} to obtain a mass of $T_{f1}$ slightly above, below, or at half the mass of the $\rho_0$. With these parameters fixed, we numerically diagonalize the mass matrices and fix the parameters $\hat{g}'$ and $\hat{g}$ by requiring the physical $W$ and $Z$ masses after diagonalization of the gauge boson mass matrix, while $y_L$ is determined by the requiring to obtain the physical top mass.
Here, we give the resulting masses, couplings, branching ratios, and cross sections for the three benchmark points in more detail. We focus on the quantities relevant to the production of the $\rho_0$ resonance, its decays, and on the decays of the top partner $T_{f1}$. The masses of the heavy resonances and top partners as well as the value of $y_L$ required to reproduce the correct SM top mass are listed in Table~\ref{tab:masses2}. The couplings relevant to $\rho_0$ production and decay are listed in Table~\ref{tab:couplings2} and \ref{tab:rho0tT2}.  

The production cross section is entirely controlled by the coupling $g_{\rho_0 qq}$ and the mass of the $\rho_0$. These values are identical in each of our benchmark points, and we obtain a production cross section of 52~fb at LHC with $\sqrt{s}=13$~TeV. Table~\ref{tab:Xsec} shows the values  of $\rho_0$ production cross section times branching ratios into the final states $\bar{t} T_{f1}$ (our signal final state), $T_{f1}\bar{T}_{f1}$ and $t\bar{t}$ (for comparison), as well as final states  $bb$, $W^+W^-$, $l_+l_-$ which are the final states yielding the strongest existing bounds on the $\rho_0$ resonance (for the model we consider).

The couplings relevant for the decay of the top partner $T_{f1}$ are given in Table \ref{tab:tTf12}. Table~\ref{tab:BRTf12} shows the resulting branching ratios of $T_{f1}$ decay channels in the three benchmark points.

\begin{table}[h!]
	\begin{center}
		\begin{tabular}{ |c | c| c | c | }
			\hline
			parameter&  Set 1 & Set 2 & Set 3  \\ \hline
			$y_L$ &0.374 &0.368 &0.371 \\		
			$m_{\rho_{0 }}$ & 2.04 & 2.04 & 2.04 \\
			$m_{T_s}$& 21.45 & 21.75& 21.75\\
			$m_{T_{f1}}$& 1.02&0.99&1.05\\
			$m_{T_{f2}}$ & 2.47& 2.64 & 2.66 \\
			$m_{X_{5/3}}$ & 1.0& 0.97 & 1.03\\
			$m_B$ & 1.04& 1.01& 1.07 \\
			\hline
		\end{tabular}
		\caption{Masses [TeV], and $y_L$ required to reproduce the SM top quark mass.}
		\label{tab:masses2}
	\end{center}
\end{table}

\begin{table}[h!]
	\begin{center}
		\begin{tabular}{ |c | c| }
			\hline
			&  Set 1 , Set 2 , Set 3  \\ \hline
			$g_{\rho_0  WW}$ & 0.12\\
			$g_{\rho_0  Zh}$ & 0.45 \\
			$g_{\rho_0  ll}$ & 0.13\\
			$g_{\rho_0  qq}$ & 0.095\\
			\hline
		\end{tabular}
		\caption{Couplings of $\rho_0$ to gauge bosons, leptons and light quarks. }
		\label{tab:couplings2}
	\end{center}
\end{table}

\begin{table}[h!]
	\begin{center}
		\begin{tabular}{ |c | c| c | c | }
			\hline
			$g_{\rho 0 \bar{X}Y} (L,R)$ &  Set 1 & Set 2 & Set 3  \\ \hline

			$ g_{\rho_0 tt}^L$ & $-3.10\times 10^{-2}$ &$-3.43\times 10^{-2}$&$-3.49\times 10^{-2}$\\
			$ g_{\rho_0 tt}^R$  &$-1.14\times 10^{-1}$&$-1.21\times 10^{-1}$&$-1.07\times 10^{-1}$\\
			$ g_{\rho_0 bb}^L$ &$-9.33\times 10^{-2}$&$-9.70\times 10^{-2}$&$-8.28\times 10^{-2}$\\
			\hline
			
		   $ g_{\rho_0 tT_{f,1}}^L$ & $7.5\times 10^{-2}$&$6.74\times 10^{-2}$&$6.94\times 10^{-2}$\\
			$ g_{\rho_0 tT_{f,1}}^R$ &$1.5$&$1.52$&$1.51$\\
		$ g_{\rho_0 tT_{f,2}}^L$ &$3.56\times 10^{-1}$&$3.61\times 10^{-1}$&$3.43\times 10^{-1}$\\
		$ g_{\rho_0 tT_{f,2}}^L$&$1.84\times 10^{-2}$&$1.85\times 10^{-2}$&$1.68\times 10^{-2}$\\
		$ g_{\rho_0 tT_{s}}^L$ & $1.44\times 10^{-2}$&$1.73\times 10^{-2}$&$1.64\times 10^{-2}$\\
			$ g_{\rho_0 tT_{s}}^R$ & $1.58\times 10^{-4}$&$1.76\times 10^{-4}$&$1.70\times 10^{-4}$\\
			\hline
			$ g_{\rho_0 X_{5/3} X_{5/3}}^L$ &$1.79$&$1.79$&$1.79$\\
				$ g_{\rho_0 X_{5/3} X_{5/3}}^R$  & $1.79$&$1.79$&$1.79$\\

			
			$ g_{\rho_0 T_{f,1} T_{f,1}}^L$ &$3.94\times 10^{-2}$&$-1.13\times 10^{-2}$&$3.43\times 10^{-2}$\\
			$ g_{\rho_0 T_{f,1} T_{f,1}}^R$ & $-1.05\times 10^{-1}$&$-1.14\times 10^{-1}$&$-9.94\times 10^{-2}$\\
		$ g_{\rho_0 T_{f,2} T_{f,2}}^L$ &$4.20\times 10^{-3}$ &$2.50\times 10^{-3}$&$2.96\times 10^{-3}$\\
			$ g_{\rho_0 T_{f,2} T_{f,2}}^R$& $4.36\times 10^{-3}$&$2.61\times 10^{-3}$&$3.09\times 10^{-3}$\\
			
			
		$ g_{\rho_0 T_{s} T_{s}}^L$& $-1.57\times 10^{-3}$&$-1.57\times 10^{-3}$&$-1.57\times 10^{-3}$\\
			$ g_{\rho_0 T_{s} T_{s}}^R$ & $-1.57\times 10^{-3}$&$-1.57\times 10^{-3}$&$-1.57\times 10^{-3}$\\

			\hline
					$ g_{\rho_0 T_{f,1} T_{s}}^L$  &  $-6.62\times 10^{-2}$&$-7.80\times 10^{-2}$&$-7.82\times 10^{-2}$\\
			$ g_{\rho_0 T_{f,1} T_{s}}^R$ & $3.11\times 10^{-3}$&$3.52\times 10^{-3}$&$3.74\times 10^{-3}$\\
				$ g_{\rho_0 T_{f,2} T_{s}}^L$ &$1.26\times 10^{-4}$ &$1.05\times 10^{-4}$&$1.17\times10^{-4}$\\
				$ g_{\rho_0 T_{f,2} T_{s}}^R$ & $-4.80\times 10^{-6}$&$-3.49\times 10^{-6}$&$-3.12\times 10^{-6}$\\
			$ g_{\rho_0 T_{f,1} T_{f,2}}^L$ &$-1.60$ &$-1.60$&$-1.61$\\
				$ g_{\rho_0 T_{f,1} T_{f,2}}^R$ &$6.70\times 10^{-1}$&$6.09 \times 10^{-1}$&$6.41\times 10^{-1}$\\
			
			\hline
			
		$ g_{\rho_0 b B}^L$ &$-5.14\times 10^{-1}$ &$-5.20\times 10^{-1}$&$-4.98\times 10^{-1}$\\
			$ g_{\rho_0 B B}^L$ & $-1.64$&$-1.63289$&$-1.65$\\
			$ g_{\rho_0 B B}^R$ &$-1.79$&$-1.79215\times 10^{-1}$&$-1.79$\\
			\hline
		\end{tabular}
		\caption{Couplings of the neutral vector resonance $\rho_0$ to 3rd generation quarks and top partners.}
		\label{tab:rho0tT2}
	\end{center}
\end{table}

\begin{table}[h!]
	\begin{center}
		\begin{tabular}{ |c | c| c | c | }
			\hline
			&  Set 1 & Set 2 & Set 3  \\ \hline
			$\lambda _{ht_R T_{f,1, L}}$ &$1.71\times10^{-2}$&$1.08\times10^{-2}$&$1.35\times10^{-2}$ \\
		$\lambda _{ht_L T_{f,1, R}}$&$-3.06\times10^{-3}$&$-2\times10^{-3}$& $-2.34\times10^{-3}$ \\
		$\lambda _{ht_R T_{f,2, L}}$ &3.67&3.62&3.81 \\
				$\lambda _{ht_L T_{f,2, R}}$ &0& 0&0 \\
		$\lambda _{ht_R T_{s, L}}$ &$-9.88\time10^{-1}$& $-9.48\times10^{-1}$&$-9.98\times10^{-1}$ \\
		$\lambda _{ht_L T_{s, R}}$ &$3.78\times10^{-1}$&$3.77\times10^{-1}$& $3.86\times 10^{-1}$ \\
				$\lambda _{htt}$  &$4.73\times10^{-2}$& $2.75\times10^{-2}$&$3.77\times10^{-2}$ \\ \hline
			$g_{ZtT_{f,1}}^L$ &$-6.57\times10^{-2}$&$-6.86\times10^{-2}$ &$-6.41 \times 10^{-2}$ \\
			$g_{ZtT_{f,1}}^R$ &$3.3\times10^{-1}$&$3.35\times10^{-1}$&$3.33\times10^{-1}$ \\
		$g_{Z T_{f,1}T_{f,1}}^L$ &$1.07\times10^{-1}$&$1.05\times10^{-1}$& $1.07\times10^{-1}$ \\
			$g_{Z T_{f,1}T_{f,1}}^R$&$9.32\times10^{-2}$&$9.13\times10^{-2}$ & $9.44\times 10^{-2}$ \\\hline
		$g_{W tT_{f,1}}^L$ &0&0 &0 \\
			$g_{W tT_{f,1}}^R$  &0&0& 0 \\
			$g_{W bT_{f,1}}^L$  &$-3.55\times10^{-3}$&$-3.68\times10^{-3}$&$-3.16\times 10^{-3}$ \\
		$g_{W bT_{f,1}}^R$  &0&0& 0 \\
			$g_{W BT_{f,1}}^L$&$3.37\times10^{-1}$&$3.37\times10^{-1}$&$3.37\times10^{-1}$ \\
			$g_{W BT_{f,1}}^R$ &$-3.42\times10^{-1}$&$-3.43\times10^{-1}$& $-3.42\times10^{-1}$ \\
			\hline
		\end{tabular}
		\caption{Couplings relevant for Top partner decay. }
		\label{tab:tTf12}
	\end{center}
\end{table}

\begin{table}[h!]
	\begin{center}
		\begin{tabular}{ |c | c| c | c | }
			\hline
		[fb]	&  Set 1 & Set 2 & Set 3  \\ \hline
			$\sigma(pp \to \rho_0) \times BR(\rho_0 \to l_+l_- )$ &0.16&0.14&0.16\\
			$\sigma(pp \to \rho_0)  \times BR(\rho_0 \to bb)$ &0.12&0.10&0.14\\
				$\sigma(pp \to \rho_0)  \times BR(\rho_0 \to W_+W_- )$ &3.55&3.26&3.59\\
					$\sigma(pp \to \rho_0)  \times BR(\rho_0 \to t t )$ &0.16&0.39&0.36\\
					$\sigma(pp \to \rho_0)  \times BR(\rho_0 \to t T_{f1} )$ &38.75&38.15&38.6\\
					$\sigma(pp \to \rho_0)  \times BR(\rho_0 \to T_{f1} T_{f1} )$ &-&0.03&-\\
			\hline
		\end{tabular}
		\caption{Production cross section at 13 TeV, for the final states whose bounds are strongest for the search channel topology of interest here. 	$\sigma(pp \to \rho_0)$ is 52 fb for set 1,  2 and 3.   }
		\label{tab:Xsec}
	\end{center}
\end{table}

\begin{table}[h!]
	\begin{center}
		\begin{tabular}{ |c | c| c | c | }
			\hline
			BR &  Set 1 & Set 2 & Set 3  \\ \hline
			&  $\Gamma/m = 1.3\times10^{-1} $ &$\Gamma/m = 1.26\times10^{-1}$ &$\Gamma/m = 1.4\times10^{-1}$ \\ \hline
			$\text{BR(}T_{\text{f1}}\text{-$>$ W b)}$ &$1.57\times10^{-4}$&$1.64\times10^{-4}$ &$1.23\times10^{-4}$\\
			$\text{BR(}T_{\text{f1}}\text{-$>$ Z t)} $& $9.9\times10^{-1}$& $9.9\times10^{-1}$ & $9.9\times10^{-1}$ \\
			$\text{BR(}T_{\text{f1}}\text{-$>$ t H)} $&$2.02\times10^{-5}$ & $8.25\times10^{-6}$&$1.17\times10^{-5}$ \\
			\hline
		\end{tabular}
		\caption{Branching ratios for the $T_{f1}$ }
		\label{tab:BRTf12}
	\end{center}
\end{table}

\bibliographystyle{JHEP-2-2}

\bibliography{VnVLQ}

\end{document}